\documentclass[useAMS,usenatbib]{mn2e}
\pdfoutput=1 
\usepackage{natbib}
\usepackage{times}
\usepackage{verbatim}
\usepackage{amsmath}
\usepackage{graphicx}
\usepackage{ulem}
\usepackage{color}
\usepackage{rotating}
\usepackage{lscape}
\usepackage{aas_macros}

\usepackage{amssymb}   
\usepackage{setspace}  

\newcommand{\etal}{{\it et~al.}}
\newcommand{\be}{\begin{equation}}
\newcommand{\ee}{\end{equation}}
\newcommand{\kms}{\,km\,s$^{-1}$}

\newcommand{\sig}{$\sigma$}
\newcommand{\lsun}{\hbox{${\rm\, L}_{\odot}$}}

\newcommand{\vd}{$\log \sigma_{0}$}
\newcommand{\sbmag}{$\langle \mu_{e} \rangle$}
\newcommand{\sblum}{$\log \langle I_{e} \rangle$}
\newcommand{\jhk}{\mbox{$JHK$}}
\newcommand{\br}{\mbox{$b_{\rm\scriptscriptstyle J}r_{\rm\scriptscriptstyle F}$}}

\title[6dFGS Fundamental Plane Data]
{The 6dF Galaxy Survey: Fundamental Plane Data}

\author[Campbell et al.]{Lachlan A. Campbell$^{1}$, 
John R. Lucey$^{2}$\thanks{Email: john.lucey@durham.ac.uk},
Matthew Colless$^{1,3}$,
D. Heath Jones$^{1,4}$,
\newauthor
Christopher M. Springob$^{1,5,6}$, 
Christina Magoulas$^{1,7}$
Robert N. Proctor$^{8}$,
\newauthor
Jeremy R. Mould$^{7,9}$,
Mike A. Read$^{10}$,
Sarah Brough$^{1}$,
Tom Jarrett$^{11,12}$,
\newauthor
Alex I. Merson$^{2,13}$,
Philip Lah$^{3}$,
Florian Beutler$^{5,14}$,
Michelle E. Cluver$^{1,11}$,
\newauthor
and Quentin A. Parker$^{1,15}$\\
$^1$Australian Astronomical Observatory, PO Box 915, North Ryde, NSW 1670, Australia\\
$^2$Department of Physics, Durham University, Durham DH1 3LE, UK\\
$^3$Research School of Astronomy \& Astrophysics, The Australian National University, Canberra, ACT 2611, Australia\\
$^4$School of Physics, Monash University, Clayton, VIC 3800, Australia\\
$^5$International Centre for Radio Astronomy Research, The University of Western Australia, Crawley, WA 6009, Australia\\
$^6$ARC Centre of Excellence for All-sky Astrophysics (CAASTRO)\\
$^7$School of Physics, University of Melbourne, Parkville, VIC 3010, Australia\\
$^8$Observat\'{o}rio Nacional, Rua Gal. Jos\'{e} Cristino 77, 20921-400 Rio de Janeiro, Brasil\\
$^9$Centre for Astrophysics \& Supercomputing, Swinburne University of Technology, PO Box 218, Hawthorn, VIC 3122, Australia\\
$^{10}$Institute for Astronomy, School of Physics and Astronomy, Royal Observatory, Blackford Hill, Edinburgh EH9 3HJ, UK\\
$^{11}$Spitzer Science Center, California Institute of Technology, Pasadena, CA 91125, USA\\
$^{12}$Astronomy Department, University of Cape Town, Private Bag X3, Rondebosch 7701, South Africa\\
$^{13}$Department of Physics and Astronomy, University College London, Gower Street, London WC1E 6BT, UK\\
$^{14}$Lawrence Berkeley National Laboratory, 1 Cyclotron Road, Berkeley, CA 94720, USA\\
$^{15}$Department of Physics \& Astronomy, Macquarie University, Sydney, NSW 2109, Australia\\
}

\begin{document}

\maketitle

\begin{abstract}
We report the 6dFGS Fundamental Plane (6dFGSv) catalogue that is used
to estimate distances and peculiar velocities for nearly 9\,000
early-type galaxies in the local (z$<$0.055) universe.  Velocity
dispersions are derived by cross-correlation from 6dF V-band spectra
with typical S/N of 12.9\,\AA$^{-1}$ for a sample of 11\,315 galaxies;
the median velocity dispersion is 163\,km\,s$^{-1}$ and the median
measurement error is 12.9\%. The photometric Fundamental Plane (FP)
parameters (effective radii and surface brightnesses) are determined
from the $JHK$ 2MASS images for 11\,102 galaxies.  Comparison of the
independent $J$- and $K$-band measurements implies that the average
uncertainty in $X_{FP}$, the combined photometric parameter that
enters the FP, is 0.013\,dex (3\%) for each band. Visual
classification of morphologies was used to select a sample of nearly
9\,000 early-type galaxies that form 6dFGSv.  This catalogue has been
used to study the effects of stellar populations on galaxy scaling
relations, to investigate the variation of the FP with environment and
galaxy morphology, to explore trends in stellar populations through,
along and across the FP, and to map and analyse the local peculiar
velocity field.
\end{abstract}


\section{Introduction}
The primary goals of the 6-degree Field Galaxy Survey
\citep[6dFGS;][]{jones04, jones05, jones09} were, first, to improve
our knowledge of the cosmography of the nearby universe over the
southern hemisphere and, secondly, to provide a sample of $\sim$10\,000
nearby early-type galaxies in order to study their intrinsic properties
and to measure peculiar velocities via the determination of Fundamental
Plane distances.

An important aspect of 6dFGS is that all observations were obtained on a
single system, the UK Schmidt Telescope (UKST) and the 6-degree Field
spectrograph \citep[6dF;][]{watson98}, resulting in a homogeneous
spectroscopic dataset. The 6dF system provided 150 optical fibres, each
with a 6.7~arcsec diameter aperture, that were deployed across a field of view of
nearly 6~degrees. 6dFGS covers the entire southern hemisphere with
$\lvert$b$\rvert$$>$10$^{\circ}$. The primary galaxy sample for the
6dFGS was selected in the K$_{s}$ band from the Two Micron All Sky
Survey \citep[2MASS;][]{jarrett00xsc}. Near-infrared (NIR) selection
means the primary sample provides a relatively unbiased (i.e.\
approximately mass-selected) survey of galaxies with old stellar
populations in the nearby universe. Over the period of the 6dFGS
observations, from 2001 to 2006, survey observations produced
approximately 137\,000 spectra. The 6dFGS database (available at
www-wfau.roe.ac.uk/6dfgs) contains these spectra along with 124\,000
galaxy redshifts and the 2MASS photometry and images. The 
median redshift is 0.05 and the redshift completeness of 6dFGS is 88\%;
full details are given in \cite{jones09}.

As well as the galaxy redshift survey, 6dFGS had two further major aims
that focussed on the early-type galaxies.

The first aim was to provide an NIR-selected spectroscopic and
photometric sample with which to study the Fundamental Plane (FP), the
scaling relation linking the velocity dispersion, effective radius and
surface brightness of early-type galaxies
\citep{dressler87,djorgovski87}. The 6dFGS subsample of early-type
galaxies with measured FP parameters has been used to investigate the
physical origins of the FP, and its implications for galaxy formation
and evolution, by exploring variations in the FP with galaxy and
environmental properties, including stellar population and morphology
\citep[see][]{proctor08,magoulas12,springob12}.

The second aim was to compile a large homogeneous peculiar velocity
catalogue derived from NIR FP distances in order to map the nearby
large-scale mass distribution for cosmological studies. Peculiar
velocities are a direct tracer of the underlying distribution of mass in
the universe, so a combined redshift and peculiar velocity survey over
the same volume can provide even better constraints on parameters of
cosmological interest than a survey of redshifts alone
\citep{burkey04,zaroubi05}. Several recent reconstructions of the
peculiar velocity field in the southern hemisphere have used the
redshifts provided by 6dFGS \citep[e.g.][]{erdogdu06,lavaux10,lavaux11}.

While the first studies of FP-based peculiar velocities used velocity
dispersions that were measured individually
\citep[e.g.][]{davies87,lynden-bell88}, the advent of multi-object fibre
systems dramatically improved the observing efficiency for measuring
velocity dispersions, particularly when studying galaxies in rich
clusters \citep[e.g.][]{colless87,lucey88,jorgensen95b}. The 6dFGS
peculiar velocity survey (6dFGSv) was designed to build on the success
of these first fibre-based studies and provide a large set of FP
peculiar velocities over the whole of the southern hemisphere. The
6dFGSv sample was selected from the brightest (and so highest
spectroscopic S/N) ellipticals, lenticulars and early-type spiral bulges
in the primary 6dFGS redshift sample over the volume out to
$\sim$16\,500\kms.

Here we present the 6dFGSv dataset. The structure of the paper is as follows. 
In Section~2 we describe in detail the procedures used to determine
velocity dispersions of 11\,315 early-type galaxies and provide a 
comparison with external catalogues. 
In Section~3 we outline the techniques we adopted to measure 
PSF-corrected FP photometric parameters (R$_e$ and
$\langle\mu_{e}\rangle$) from the 2MASS image tiles and demonstrate
that these new NIR measurements are in excellent agreement with values
previously derived from NIR and optical data.
In Section~4 we describe the visual classification of the 6dFGSv galaxy
morphologies, which is used to restrict the sample to those objects
likely to yield reliable FP results. The construction of the 6dFGSv
catalogue that is used in our FP studies is reported in Section~5.
Finally, we present some brief conclusions and highlight ongoing studies 
in Section~6.

\section{Spectroscopic Measurements}
\subsection{6dF spectra and sample selection}

The main characteristics of the 6dFGS, including the 6-degree Field
(6dF) instrument, target catalogue construction, the allocation of
fields and fibres to targets, data reduction, and procedures for
redshift determination, are reported in \cite{jones04}. Briefly, 6dF
uses 6.7\,arcsec diameter fibres to observe 150 spectra simultaneously
covering two spectral ranges; the 'V-band' data cover 3900--5600\,\AA\
and the `R-band' data cover 5400--7500\,\AA. When estimating redshifts,
these two bands were spliced together to maximize the signal. However,
to avoid problems with discontinuities and dispersion changes only the
V-band spectra were used to estimate velocity dispersions. The pixel
scale was 1.64\,\AA\ and the average 1\sig\ spectral resolution was
$\sim$140\kms.

Not all the $\sim$136\,000 galaxy spectra of the full 6dFGS redshift
survey are suitable for velocity dispersion measurements, and we applied
a number of criteria in order to define the 6dFGSv sample. We only
selected galaxies with the higher quality redshift flags, $Q=3$--5, see
\cite{jones04}. This ensured the selected spectra had reliable redshifts
before their use for velocity dispersion information. At redshifts
greater than 16\,500\kms\ the important strong Mg$b$ feature starts to
be affected by the red wavelength cut-off of the 6dF V-band spectra;
hence we excluded galaxies with redshifts greater than 16\,500\kms.
Further selection was made by spectral classification, based on the
best-match template spectra identified in the redshift estimation.
This was initially based on the R-value returned from our cross-correlation 
analysis (see Section 2.2) which reflected how closely the galaxy 
spectrum matched that of our K-giant star templates.  
We found that spectra with a R-value of 8 had a
velocity dispersion uncertainty of $\sim$0.12 dex.
Hence as a practical choice we used a R $>$ 8 criteria to select spectra 
that gave sufficiently reliable velocity dispersion measurements 
for our study. 
These various selection criteria restricted the total sample for
velocity dispersion measurements to $\sim$18\,000 galaxies.

While the above filtering process excluded spectra with strong emission lines, 
about $\sim$690 spectra had weak emission features.
All spectra were visually checked and poorly subtracted
sky-lines, foreground H$\beta$ emission lines and weak galaxy emission lines
(e.g.\ H$\beta$, [OIII]\,4959/5007) were removed by local interpolation.
We tested the sensitivity of our interpolation procedure by running an 
alternative pixel-based analysis in which the pixels that were previously
modified by the interpolation were flagged as invalid.
The derived velocity dispersions changed, on average, by only 1 km\,s$^{-1}$;
the rms scatter of the difference was 15 km\,s$^{-1}$. 
On average, the removal of the weak emission features by the local interpolation
technique increased the R-value by 3 and changed the derived velocity dispersions by
--7 km\,s$^{-1}$ with an rms scatter of 17 km\,s$^{-1}$.

\subsection{Velocity dispersion measurements}

Our analysis of the 6dF V-band spectra was undertaken in IRAF using the
RV package. The {\tt fxcor} routine, which implements the method of
from \cite{tonry79}, 
hereafter TD79, was used for the velocity dispersion
determination. Spectra of a set of template stars ranging in stellar 
type from G8 to K2 giants were observed with 6dF for the analysis
(see Table~\ref{tab:tspectemplates}).

\begin{table}
\caption{6dFGSv stellar templates used with {\tt fxcor}.}
\label{tab:tspectemplates}
\begin{tabular}{llrrrc}
 Name  & Type & Velocity & Plate & Fibre & FWHM\\
 & &  \kms & & & \AA\\
 \hline
HD\,223679   & G8\,III   &   +4 & 2 &  62 & 5.672 \\
HD\,813      & G8/K0\,IV &  +24 & 2 & 109 & 5.211\\
HD\,225166   & G8/K0\,IV & +123 & 1 &  86 & 5.365\\
CD-45\,15239 & K0        &  +39 & 2 & 101 & 5.107\\
HD\,214      & K1\,III   &  +63 & 1 & 130 & 5.459\\
HD\,224420   & K1\,III   &  +66 & 1 & 116 & 5.212\\
HD\,321      & K2\,III   &  +39 & 2 & 137 & 5.154\\
\hline 
\end{tabular}
\end{table}

To prepare for cross-correlation, the spectra were shifted to a
rest-frame wavelength based on the redshift calculated in {\tt runz}
\citep{colless01, saunders04} 
and rebinned to a log-wavelength scale. Spectra were
truncated to have a minimum wavelength of 4000\,\AA, and a maximum limit
depending upon the redshift of the galaxy; this ranged from 5570\,\AA\
at $z$=0 to 5280\,\AA\ at $z$=0.055. Cropping the spectra in this way
gave the maximal coverage of desired spectral features such as Mg and Fe
lines, while excluding problematic features such as the night sky line
residuals of [OI]\,5577 and intrinsically broad Ca~H~\&~K spectral
lines. The spectral continuum was then subtracted and 5\% of the pixels
at each end of the spectra were apodized to prevent ringing in the
correlation. The spectra were zero-padded and converted to Fourier
space, then filtered with a ramp filter to remove any high-frequency
noise or low-frequency power due to continuum residuals and the
apodisation.

The galaxy spectra were correlated with the templates, and the largest
peak in Fourier space was fit over a range of eleven pixels using a
Gaussian function to determine the centre and FWHM. The FWHM was
converted to an estimated galaxy velocity dispersion using the
empirically established response, for each template, of the
cross-correlation peak to velocity broadening. The final velocity
dispersion was the average of all templates, weighted by the strength of
the cross-correlation, $R$.

\begin{figure}
\includegraphics[width=8cm]{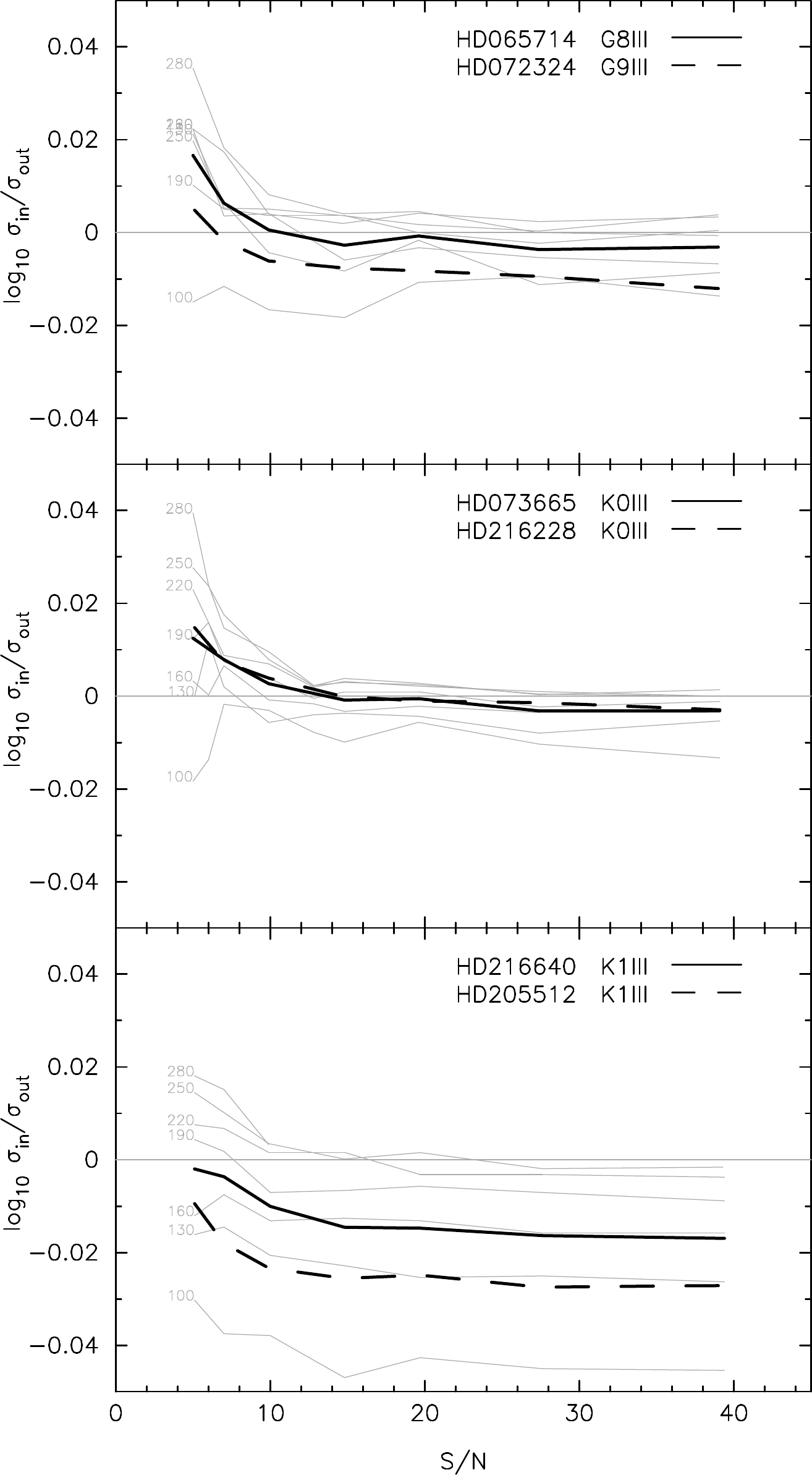}
\caption{
Our {\tt fxcor}-based pipeline systematics.
The log ratio of the input and the recovered velocity dispersions
is plotted against S/N. The input velocity dispersions tested
were 100, 130, 160, 190, 220, 250 and 280 km\,s$^{-1}$. The black lines 
are the average values. The grey lines show the individual velocity 
dispersions; for clarity, these are only plotted for the 
first star listed in each panel.}
\label{boot_sys}
\end{figure}

As an independent test of our velocity dispersion measurement procedure 
we used the spectra from the MILES library \citep{sanchez06} which have a
FWHM spectral resolution of 2.54\,\AA, i.e. a factor of two lower than
the 6dF dataset. We selected six stars with spectral types ranging from G8III 
to K1III. These were broadened to a range of velocity dispersions,
degraded to a range of S/Ns and matched to the spectral range and 
resolution characteristics of the 6dF dataset. For each 
input velocity dispersion and S/N we produced a set of $\sim$1000 mock spectra 
and these were processed with our pipeline using the 
templates listed in Table~\ref{tab:tspectemplates}. 
The systematic differences between the input velocity dispersions 
and those recovered are presented in 
Figure~\ref{boot_sys}. The systematic biases are typically less than 
0.02 dex. At lowest S/N the biases increase to 0.04 dex.
Our {\tt fxcor}-based pipeline best recovers the input velocity dispersions from the 
spectra constructed from the two K0III stars as this is the average spectral type 
of our templates.

\begin{figure}
\centering
\includegraphics[width=8cm]{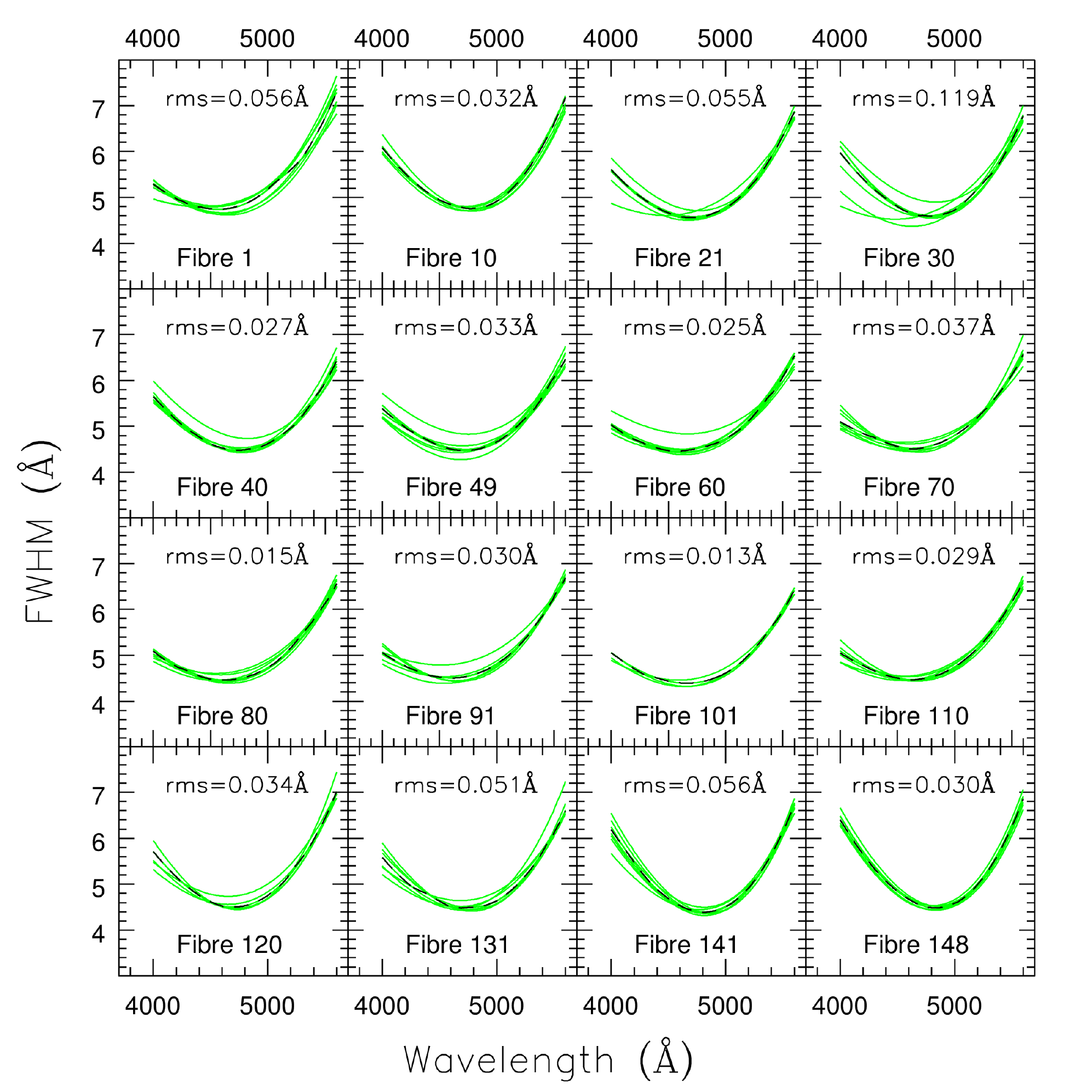}
\caption{Instrumental spectral FWHM of V-band spectra for a subset of
  6dF fibres. The instrumental FWHM were measured from multiple HgCdHe
  arc spectra observations and fitted with a third-order polynomial. The
  dashed line shows the median resolution as a function of wavelength. A
  FWHM of 5\,\AA\ corresponds to a 1\sig\ resolution of 127\kms.}
\label{fibrevar}
\end{figure}

The 6dF instrument, like many earlier multi-fibre spectrograph systems
(e.g.\ WYFFOS, see \citealt{moore02}), has an instrumental profile that
varies non-linearly both along the spectrum from a single fibre and from
fibre-to-fibre (see Figure~\ref{fibrevar}). Also, over the six years of
the 6dF survey, the resolution showed variations due to causes such as
spectrograph focus and fibre degradation. The average 6dF spectral
resolution was 140\kms\ with 95\% of fibres having mid-wavelength
resolutions between 128\kms\ and 150\kms. Hence there were, in some
cases, significant differences in resolution between the template and
object spectra. It is implicit in the TD79 method that the instrumental
resolution of the template spectrum and the galaxy spectrum be
identical, so that
\begin{equation}
\mu^2=\sigma^2+2\tau^2,
\end{equation}
where $\mu$ is the width of the cross-correlation peak, $\tau$ is the
dispersion due to instrumental resolution, and \sig\ is the dispersion
due to velocity broadening. The fact that this was not always the case
would therefore result, unless corrected, in biased velocity dispersion
estimates.

Any difference in resolution between template and object spectra
($\Delta\tau$) is simply added in quadrature to the cross-correlation
peak width. However, prior to the cross-correlation the spectra are
filtered and, as seen in Figure~1 of TD79, this changes the relationship
between the width of lines in the spectra, $\tau$, and the width of the
peak, $\mu$. To properly correct for differences in resolution we
measured the change in $\mu$ as a function of $\Delta\tau$.
Auto-correlation of a stellar template was performed over a range of
$\tau$ values equivalent to adding in quadrature template and object
spectra with differences in resolution of between $-$40\kms\ and
+40\kms. The $\mu$ values responded in a linear fashion to changes in
$\tau$, and were essentially independent of the stellar template and the
wavelength range of the spectra. Based on these results, a look-up table
was generated giving the fits as a function of $\Delta\tau$, from which
a corrected cross-correlation peak width $\mu$ could be obtained. A
zeroth order estimation of the resolution of the template and object
spectra, centred roughly around 5200\,\AA\ (the major source of signal
in the Fourier cross-correlation comes from the Mg and Fe lines in
early-type galaxies), gave $\Delta\tau$ for each observation, and the
corresponding $\mu$ was then converted to a resolution-corrected
velocity dispersion using the normal template calibration curve.

\begin{figure}
\centering
\includegraphics[width=8cm]{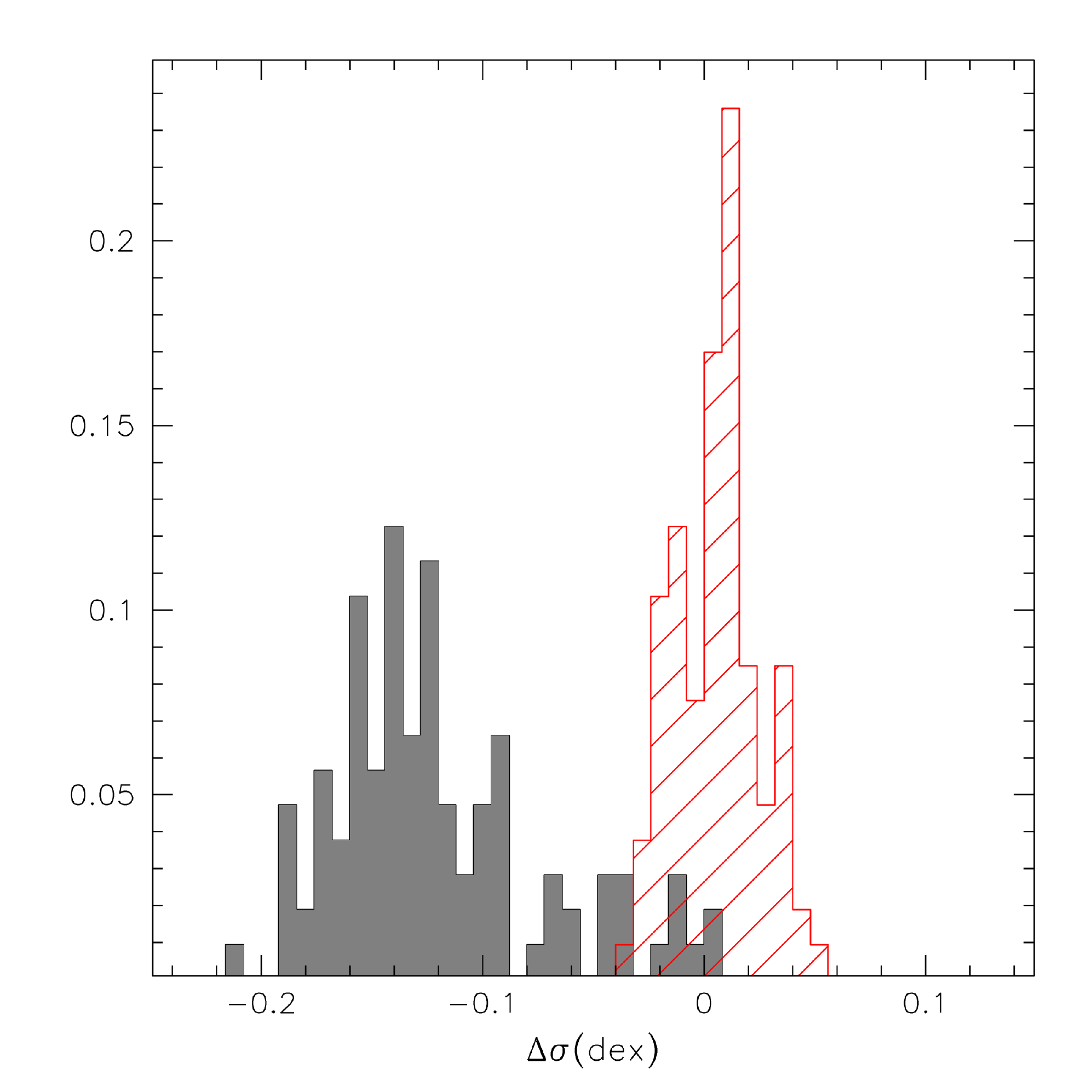}
\caption{Skyflat velocity dispersions for 6dF fibre \#4 before and after
  applying a resolution correction. The distributions of the differences
  between skyflat velocity dispersions before (the shaded histogram) and
  after (the hatched histogram) applying a resolution correction for the
  case of an applied broadening of 150\kms\ and a template with a
  spectral resolution of 166\kms. This is the case with the poorest
  resolution and hence the largest correction.}
\label{frescorrectionsky}
\end{figure}

Twilight exposures (skyflats), which result in high S/N solar spectra
being observed in each fibre, provided a useful test of our resolution
correction procedure. The source spectrum is in each fibre is identical,
but the different fibres and positions on the camera mean the resolution
naturally varies. We selected three fibre spectra with spectral
resolutions of 126, 140 and 166\kms\ to act as the spectral templates.
All the twilight spectra were then broadened by 100, 150 and 200\kms\ to
simulate the galaxy observations. We then applied our resolution
correction procedure and confirmed that the input broadening was
recovered. Figure~\ref{frescorrectionsky} shows the results of using the
166\kms\ resolution fibre as the template with the twilight sky spectra
broadened by 150\kms, and demonstrates the validity of our procedure.

As a further independent test of our resolution correction, we examined
120 galaxies that also had velocity dispersion measurements from SDSS.
We divided these galaxies into sub-samples with lower and higher 6dF
resolution ($\tau<135$\kms\ and $\tau>145$\kms). The comparison
confirmed the existence of the resolution bias, and validated the
correction procedure. It also demonstrated that the correction
introduced no appreciable increase in scatter ($\leq$10\kms).

\begin{figure}
\centering
\includegraphics[width=6cm,angle=270]{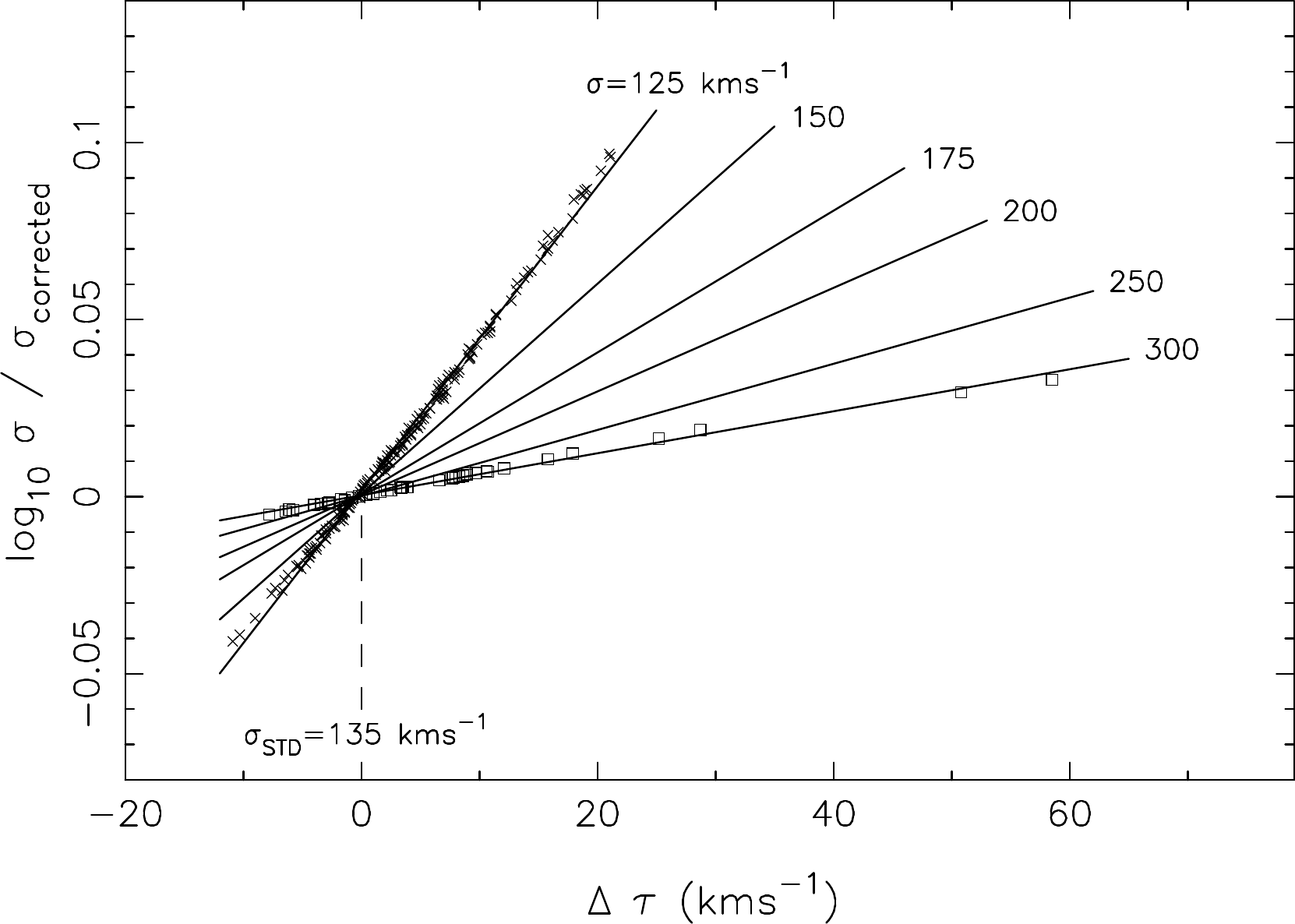}
\caption[Corrections to $\sigma$, as a function of the difference in
  resolution between template and object
  spectra]  
{\label{frescorrection6dfgsv}The scale of these corrections,
   as a function of the difference in resolution between
  template and object spectra ($\Delta\tau$), for the velocity
  dispersion sample. The lines show the average correction applied for
  objects with a \sig~of 125, 150, 175, 200, 250, and 300 \kms. For
  clarity, the actual corrections applied are only shown for
  individual galaxies with $\sigma\sim125$ \kms ~(crosses) and
  $\sigma\sim300$ \kms ~(squares).}
\end{figure}

Consequently, the resolution correction using this method 
were applied to all the 6dFGS velocity dispersion measurements. 
Figure \ref{frescorrection6dfgsv} shows the scale of these
corrections, as a function of the difference in
resolution between template and object spectra ($\Delta\tau$), for
the velocity dispersion sample. The figure clearly illustrates how
the correction had a much more significant impact for lower velocity
dispersion measurements. Moreover, the correction is simple,
well-behaved at all velocity dispersions, and straightforward, supporting the
adoption of the correction procedure.

\begin{figure}
\includegraphics[width=8cm]{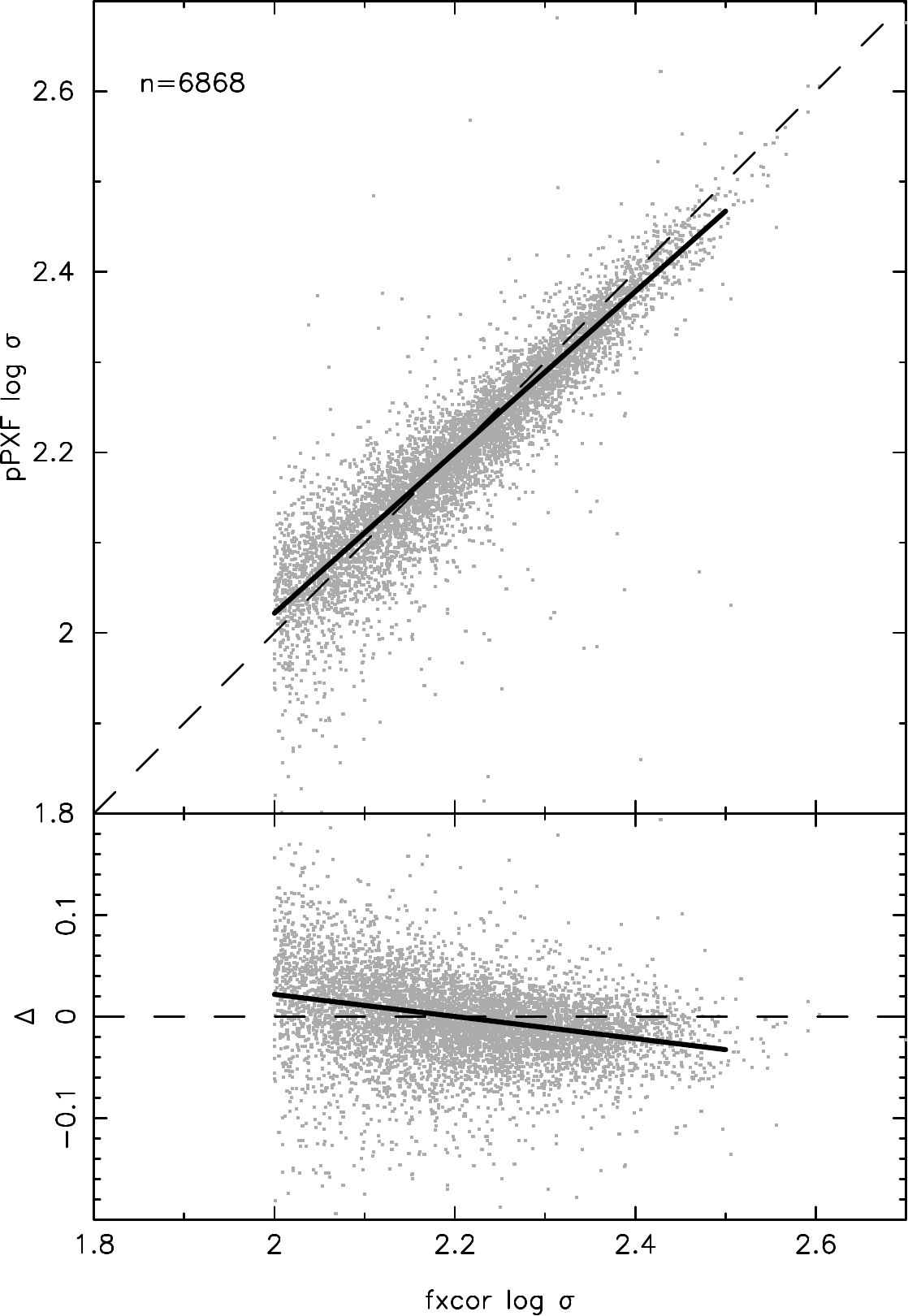}
\caption{
Comparison of {\tt fxcor}- and pPXF-based velocity dispersion measurements.
The solid line indicates the best fit linear relation between the two 
sets of measurements. This has the form: 
$\log{\sigma{\rm (pPXF)}}$\,=\,0.891\,$\times$\,$\log{\sigma{\rm(fxcor)}}$\,+\,0.239 .
}
\label{fxcor_ppxf}
\end{figure}

While many previous FP peculiar velocity surveys have used 
the cross-correlation technique as implemented in {\tt fxcor} 
to measure velocity dispersions, e.g. the ENEAR survey 
\citep{wegner03} and the NOAO Fundamental Plane survey (\citealt{smith04}; NFPS), 
recent studies of galaxy dynamics have mainly used pixel-based fitting 
techniques. In order to directly assess the difference between these two 
widely used techniques we have measured velocity dispersions with pPXF
\citep{cappellari04} for a subset of our spectra. 
We selected the four highest resolution 
templates (HD\,813, CD-45\,15239, HD\,224420, HD\,321) 
and the galaxy spectra that were observed with 
instrumental resolutions that were greater than that of the four 
templates. Extra instrumental (gaussian) broadening 
was added to the templates to match the individual resolutions of the galaxy 
spectra and the velocity dispersions were determined using pPXF.
There is good agreement between the two techniques with only
a moderate systematic difference (see Figure~\ref{fxcor_ppxf}).
The observed systematic offsets at velocity dispersions of 2.00 and 2.50\,dex 
are relatively small, i.e. +0.022\,dex and {--0.032}\,dex, respectively. 
These differences between the methods are comparable to the
systematic variations that can result from template mismatch (see
Fig.1). We conclude that both fxcor and pPXF are limited by systematic
uncertainties from template mismatch and spectral filtering to a
precision of about 0.02 dex. 
As shown below this is substantially less than the 
typical random error for the 6dFGS velocity dispersions.

We note that any systematic trend in velocity dispersion 
measurements will influence the derived parameters
of the FP, notably the coefficient associated with the $\log{\sigma}$ term.
Clearly this is a potential source of bias and is particularly 
relevant when combining datasets from different sources. 
Fortunately there is now a large number of early-type galaxies with 
velocity dispersion measurements from more than one source
(see Section 2.5) which allow this bias to be fully investigated.

\subsection{Velocity dispersion uncertainties}
\label{Errors}

Measurement errors in velocity dispersions are due to a number of
factors: template--galaxy mismatch, noise in the spectra, non-optimal
filtering, and calibration errors. Errors due to template mismatch were
minimized by using appropriate templates, which in the case of
early-type galaxies means late G-type and early K-type stars. Filter
values were optimized using a range of stellar templates. Calibration
errors depend upon the accuracy of the function used to fit the template
calibration curve, in this case a fifth order polynomial. Twilight flats
were used to test the scale of calibration errors, and these were found
to be negligible, as they did not exceed 0.005\,dex for velocity
dispersions larger than $\sim$100\kms.

\begin{figure}
\centering
\includegraphics[width=8cm]{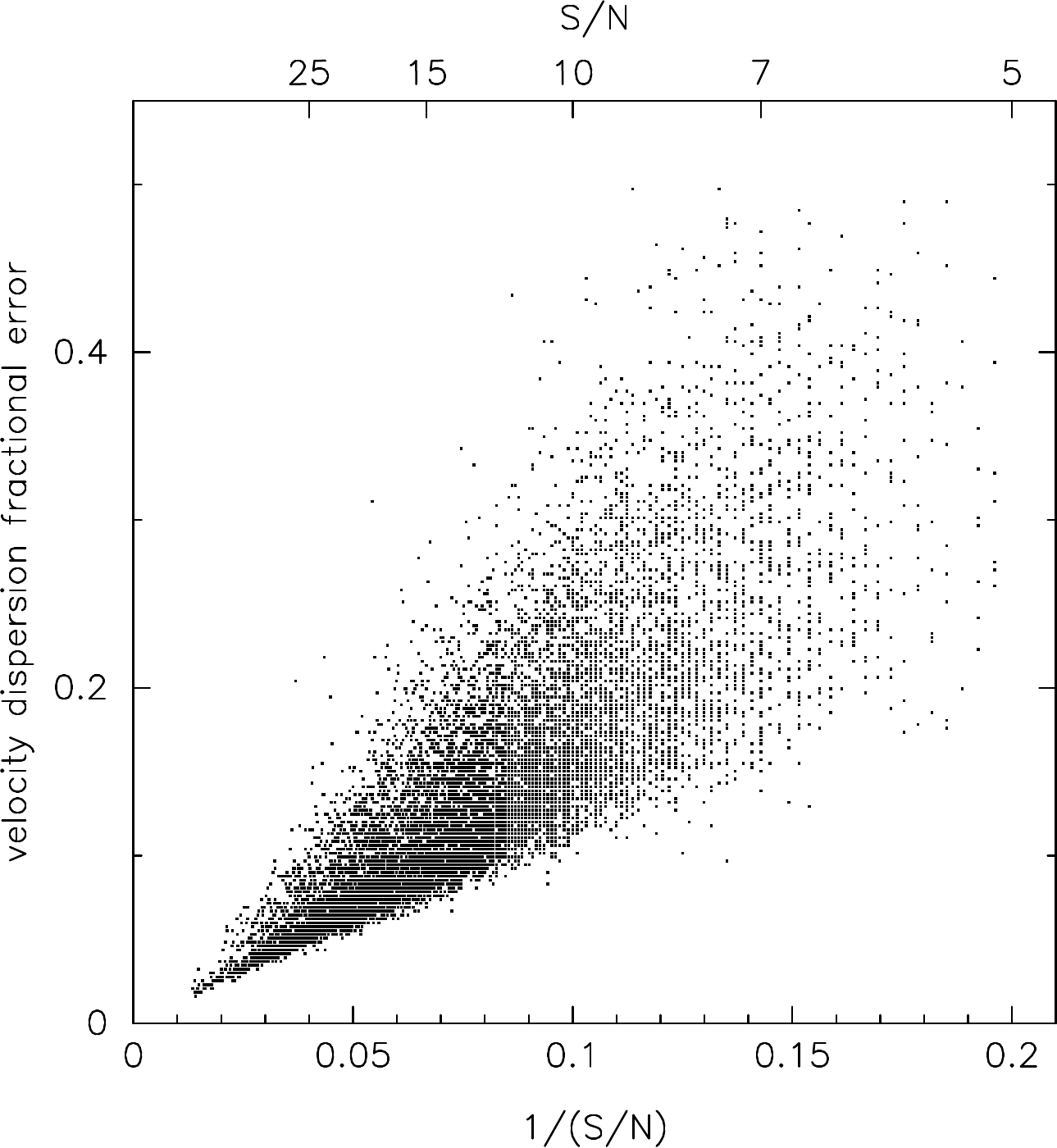}
\caption{
Velocity dispersion fractional error as a function of inverse S/N.
0.8, 4.7 and 19.8 percent of the 6dFGSv sample have fractional 
errors greater than 0.4, 0.3 and 0.2 respectively.}
\label{boot_vs_s2n}
\end{figure}

The statistical uncertainties of our velocity dispersion measurements
were determined via a bootstrap technique. Along with the galaxy
spectrum, the 6dF data reduction pipeline also 
provides an associated variance spectrum.
We used the galaxy and variance spectra to create many ($\sim$900) bootstrap 
realisations of the original spectrum and for each we measured the velocity dispersion. 
We adopted the standard deviation of these measurements 
as the statistical error of the 
velocity dispersion.
This procedure was tested by using synthetic 
spectra that mimicked the characteristics of the 6dF spectra, 
i.e. spectral coverage, velocity dispersion and S/N range, and 
was found to give reliable estimates of the measurement errors.
For the lowest S/N spectra modelled, i.e. S/N\,=\,5, 
this technique returns values for the errors that
are corrected, on average, at the level of a few ($\sim$5) percent.
Figure~\ref{boot_vs_s2n} illustrates how the fraction error depends on the inverse S/N.

\subsection{The 6dFGSv velocity dispersion catalogue }

To construct our final 6dFGSv velocity dispersion sample, several
criteria were applied. As previously noted, we excluded galaxies with
$z>0.055$, where the strong Mg$b$ feature is affected by the red limit
of the V-band spectral coverage. In order to ensure a reliable set of
measurements, we only included galaxies that had spectral measurements
with S/N\,$>$\,5, $R$\,$>$\,8
and \sig\,$>$\,100\kms. These criteria resulted in a
velocity dispersion sample with 11\,520 measurements;  185 galaxies have
two measurements, ten have three measurements. 

The 6dFGSv velocity dispersion catalogue is presented in
Table~\ref{tab:sigma_data} for an example set of galaxies;
the velocity dispersion given in this table are not aperture corrected.
The full
sample has a median velocity dispersion of 163\kms, a median S/N of
12.9\,\AA$^{-1}$, and a median velocity dispersion error of 12.9\%.
 
\begin{table*}
  \caption{6dFGSv Velocity Dispersion Catalogue. 
    The columns are as follows:
    (1) 6dFGSv unique reference number;
    (2) 6dFGS input catalogue standard galaxy name; 
    (3) 2MASS Extended Source Catalogue designation;
    (4) Observation Modified Julian Date;
    (5) Heliocentric redshift; 
    (6) Signal-to-noise ratio per \AA~measured from a region between 4900-5000\,\AA;
    (7) TD79 R-value in the Fourier cross-correlation;
    (8) Velocity dispersion in \kms ~(6.7 arcsec diameter fibre aperture)
corrected for instrumental resolution but not corrected for aperture size;
    (9) Log error in the velocity dispersion, i.e. $\epsilon_{\log\sigma} \equiv \delta\log\sigma$.
    The full version of this table has 11\,315 galaxies (185 galaxies have
    two measurements, ten have three measurements; all are listed, so this table has 11\,520
    data lines) and is provided in the online Supporting Information.}
\label{tab:sigma_data}
\begin{tabular}{cccccrrcc}
\hline \hline
\multicolumn{1}{c}{\#} &  
\multicolumn{1}{c}{6dFGS ID} & 
\multicolumn{1}{c}{2MASS XSC name}  &
\multicolumn{1}{c}{MJD} &
\multicolumn{1}{c}{z$_{helio}$} & 
\multicolumn{1}{c}{S/N} & 
\multicolumn{1}{c}{TDR} & 
\multicolumn{1}{c}{$\sigma$} &
\multicolumn{1}{c}{$\epsilon_{\log\sigma}$} \\
\multicolumn{1}{c}{(1)} &  
\multicolumn{1}{c}{(2)} & 
\multicolumn{1}{c}{(3)} &
\multicolumn{1}{c}{(4)} &
\multicolumn{1}{c}{(5)} & 
\multicolumn{1}{c}{(6)} &
\multicolumn{1}{c}{(7)} & 
\multicolumn{1}{c}{(8)} &  
\multicolumn{1}{c}{(9)} \\
\hline
00001 & g0000144-765225 & 2MASXJ00001440-7652248 & 53614.60 & 0.0533 & 13.2 & 14.0 & 129.0 & 0.073\\
00002 & g0000222-013746 & 2MASXJ00002213-0137463 & 52962.44 & 0.0383 & 24.1 & 18.4 & 199.1 & 0.023\\
00003 & g0000235-065610 & 2MASXJ00002348-0656103 & 52966.46 & 0.0376 & 15.2 & 16.6 & 156.2 & 0.045\\
00004 & g0000251-260240 & 2MASXJ00002509-2602401 & 52846.78 & 0.0508 & 13.2 & 13.0 & 218.5 & 0.040\\
00005 & g0000356-014547 & 2MASXJ00003564-0145472 & 52962.44 & 0.0244 & 14.0 & 16.3 & 133.4 & 0.066\\
00006 & g0000358-403432 & 2MASXJ00003574-4034323 & 52881.58 & 0.0500 & 9.6  & 8.9  & 143.7 & 0.093\\
00007 & g0000428-721715 & 2MASXJ00004283-7217148 & 53618.50 & 0.0348 & 13.9 & 19.0 & 151.2 & 0.049\\
00008 & g0000459-815803 & 2MASXJ00004596-8158024 & 53646.46 & 0.0422 & 14.9 & 9.6  & 200.9 & 0.058\\
00009 & g0000482-551119 & 2MASXJ00004816-5511192 & 52969.51 & 0.0323 & 11.1 & 14.4 & 135.4 & 0.081\\
00010 & g0000523-355037 & 2MASXJ00005234-3550370 & 52872.63 & 0.0520 & 14.9 & 10.2 & 242.3 & 0.048\\
00011 & g0000532-355911 & 2MASXJ00005317-3559104 & 52872.63 & 0.0500 & 15.6 & 13.3 & 207.0 & 0.040\\
00012 & g0000558-255421 & 2MASXJ00005579-2554210 & 52846.78 & 0.0505 & 10.3 & 15.4 & 188.4 & 0.042\\
00012 & g0000558-255421 & 2MASXJ00005579-2554210 & 52880.53 & 0.0505 & 15.6 & 19.5 & 149.9 & 0.038\\
\hline
\end{tabular}
\end{table*}

\subsection{External Comparisons}
\begin{figure}
\centering
\includegraphics[width=8cm]{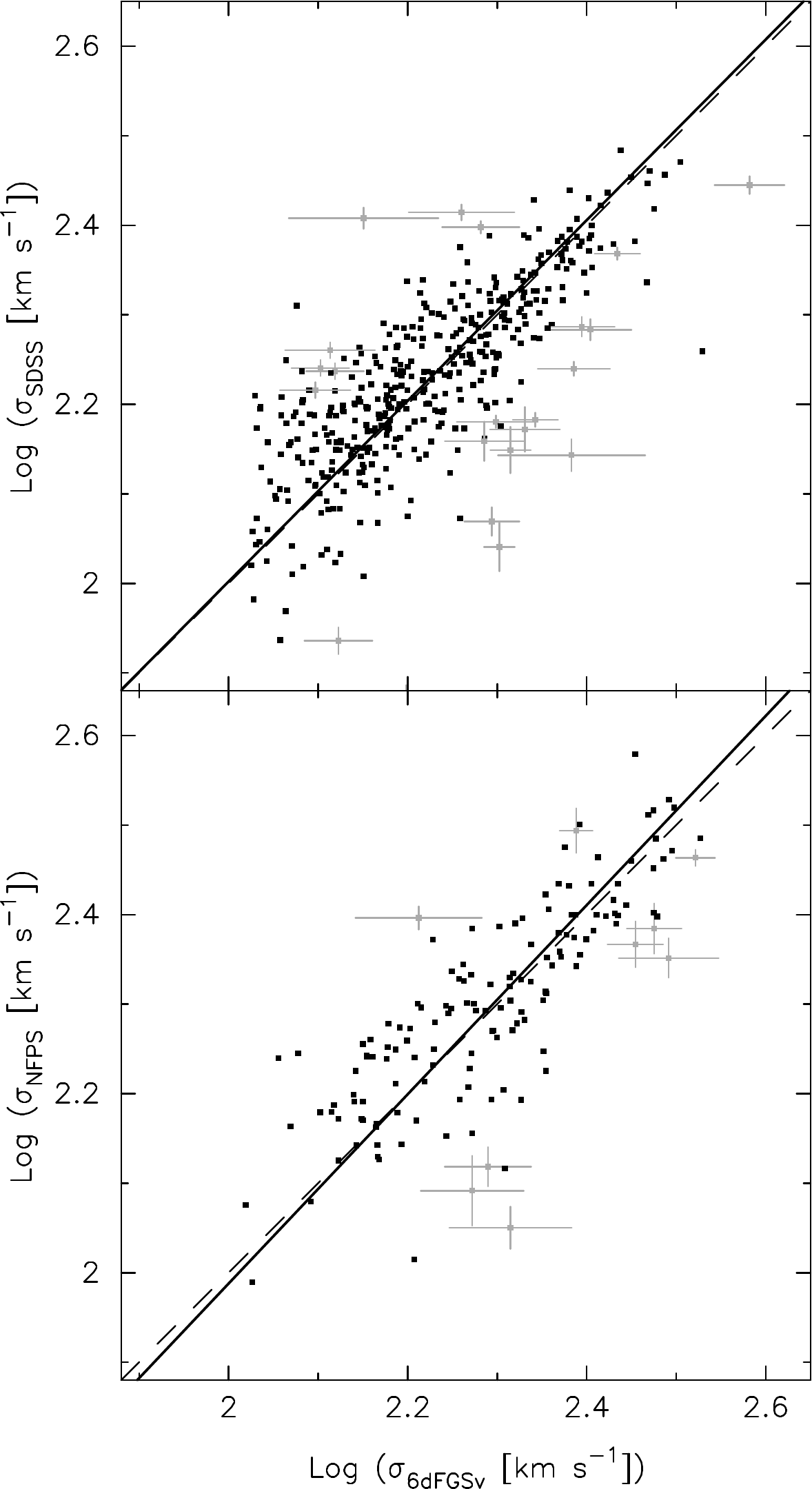}
\caption{ 
Comparisons of 6dFGSv velocity dispersion measurements with those
  from SDSS (top panel) and NFPS (lower panel). The dashed line
  shows the one-to-one relationship for reference.
  The solid line shows the best
  linear fit, calculated as described in Section 2.5.
  The black squares show galaxies included in the fit.   
  The grey squares show objects that were excluded from the fit.
  For clarity only error bars on the excluded data points are
  displayed.
}
\label{fextsigmasdssnfps}
\end{figure}

There are a sizable number of galaxies in the 6dFGSv sample that have
dispersion velocity measurements from previous large studies, i.e.
the Streaming Motions of Abell Clusters survey (\citealt{hudson01};
SMAC); the ENEAR redshift-distance survey \citep{wegner03}; the NOAO
Fundamental Plane survey (\citealt{smith04}; NFPS); and the Sloan
Digital Sky Survey DR8 (\citealt{aihara11}; SDSS).  The comparison of
the 6dFGSv velocity dispersions with those values reported by SDSS and
NFPS are presented in Figure~\ref{fextsigmasdssnfps}.  The best-fit
relation between these independent measurements was calculated by
minimizing the $\chi^2$ sum
\begin{equation}
( a\,\sigma_x\,+\,b - \sigma_y\,)^{2}\,/\, ( \epsilon(\sigma_x)^2 + \epsilon(\sigma_y)^2 )
\label{eq:sigma_comp}
\end{equation}
where $\sigma_x$ and $\epsilon(\sigma_x)$ refer to the 6dFGSv
measurements and their errors, and $\sigma_y$ and $\epsilon(\sigma_y)$
refer to the external dataset.  Data points with $\chi^2$\,$>$\,6 were
excluded from the fit.  In Table~\ref{tab:external_sigma} we summarize
the observed relations between 6dFGSv velocity dispersions and those
from the four external sources. All comparisons are consistent with
unit slope, and hence, within the typical slope uncertainty of 0.035
dex, there is no evidence of systematic bias. 
Also listed in Table~\ref{tab:external_sigma} is the median offsets
between the velocity dispersion measurements of these four external 
sources and 6dFGSv. While the only significant offset is that for the
6dFGSv\,--\,ENEAR comparison, i.e. $\Delta$ log $\sigma$ = 0.021 $\pm$ 0.005, 
this difference is similar in size to
that found when inter-comparing $\sigma$ datasets
(see \citealt{hudson01}, Table 2).
A detailed inter-comparison of the various velocity dispersion datasets, 
including the four considered here, will be presented in future work 
where a standardized all-sky velocity dispersion catalogue optimized for FP
studies will be constructed.

\begin{table*}
\caption{Comparison of 6dFGSv velocity dispersion measurements with external sources.
For this comparison
aperture corrections to the velocity dispersions have been applied, see Section 5.1.
$n$ is the number of objects common between the dataset and 6dFGSv.  
$n_{\rm FIT}$ is the
number of objects remaining after the $\chi^2$\,$>$\,6 clipping has been applied.
$a$ and $b$ are defined in Equation~\ref{eq:sigma_comp}. 
$\Delta$ log $\sigma$ is the median offset (6dF -- Source) between the two systems after clipping.
$rms$ is the scatter of the difference between the two system after clipping.
}
\label{tab:external_sigma}
\begin{tabular}{lrrllccc}
Source & $n~~$ & $n_{\rm FIT}$ & ~~~~~~~~$a$ & ~~~~~~~~$b$ & $\Delta$ log $\sigma$ & rms & $\chi^2$\\
 \hline
SDSS  &  419 & 397 & 1.009 $\pm$ 0.025 & --0.016 $\pm$  0.057 &  0.008  $\pm$ 0.004 & 0.057 & 0.984 \\
NPFS  &  146 & 137 & 1.056 $\pm$ 0.045  & --0.126 $\pm$ 0.107 &  0.010  $\pm$ 0.006 & 0.058 & 1.060 \\
SMAC  &  ~99 &  ~93 & 0.931 $\pm$ 0.035 & +0.167  $\pm$ 0.081 &  0.005  $\pm$ 0.006 & 0.045 & 1.003\\
ENEAR &  174 & 155 & 0.985 $\pm$ 0.030  &  +0.057 $\pm$ 0.068 &  0.021  $\pm$ 0.005 & 0.049 & 1.351\\
\hline 
\end{tabular}
\end{table*}

\section{Photometric Measurements}
\subsection{PSF-corrected effective radii and surface brightnesses}
The Fundamental Plane (FP) is the relationship between two photometric
parameters, a characteristic radius and the mean surface brightness
within that radius, and the central velocity dispersion. The key step in
determining a galaxy's photometric parameters is the measurement of the
total flux. Early work (e.g.\ \citealt{burstein87}) derived the total
magnitude $m_{T}$ through a $r^{1/4}$ growth curve fit to photoelectric
aperture photometry, and measured the effective (half-light) radius
$R_e$ by interpolation. The mean surface brightness 
$\langle \mu_e \rangle$ within $R_e$ was determined via

\begin{equation}
\langle \mu_e \rangle = m_T + 2.5\log(2 \pi R_e^2) ~. 
\end{equation}

For simplicity, early FP studies using CCD imaging data mimicked this
approach (e.g.\ \citealt{lucey88}, \citealt{lucey91}) with the addition
of an empirical correction for the PSF blurring. Later work using CCD
imaging data undertook more sophisticated analysis to determine the
total magnitude. For example, \citet{jorgensen95a} fitted ellipses to
the surface photometry and used model fitting ($r^{1/4}$)
of the growth curves of the photometry.
\cite{bernardi03} used the $r^{1/4}$ (or exponential)
`model' magnitudes from the SDSS image processing software which are
derived from PSF-convolved two-dimensional surface fitting. Alternative
approaches used to determine $R_e$ include fitting seeing-convolved
S\'{e}rsic models to circularized profiles or major and minor axis
profiles \citep{saglia97,donofrio08} or full two-dimensional surface
brightness fitting of a S\'{e}rsic model convolved with a point spread
function \citep{gargiulo09} or S\'{e}rsic plus exponential disk models
(e.g.\ \citealt{fernandez11}).

For the 6dFGSv sample we analysed the imaging data from the 2MASS survey
(Skrutskie \etal\ 1997) to derive the FP photometric parameters. The key
advantage of the 2MASS data is the near-complete all-sky coverage and
the excellent photometric uniformity. Furthermore, the online NASA/IPAC
Infrared Science Archive provides straightforward access to the 2MASS
image tiles, which allows the measurement of additional phototometric
parameters not provided by the 2MASS Extended Source Catalog (XSC;
Jarrett \etal\ 2000).

\begin{figure}
\centering
\includegraphics[width=0.45\textwidth]{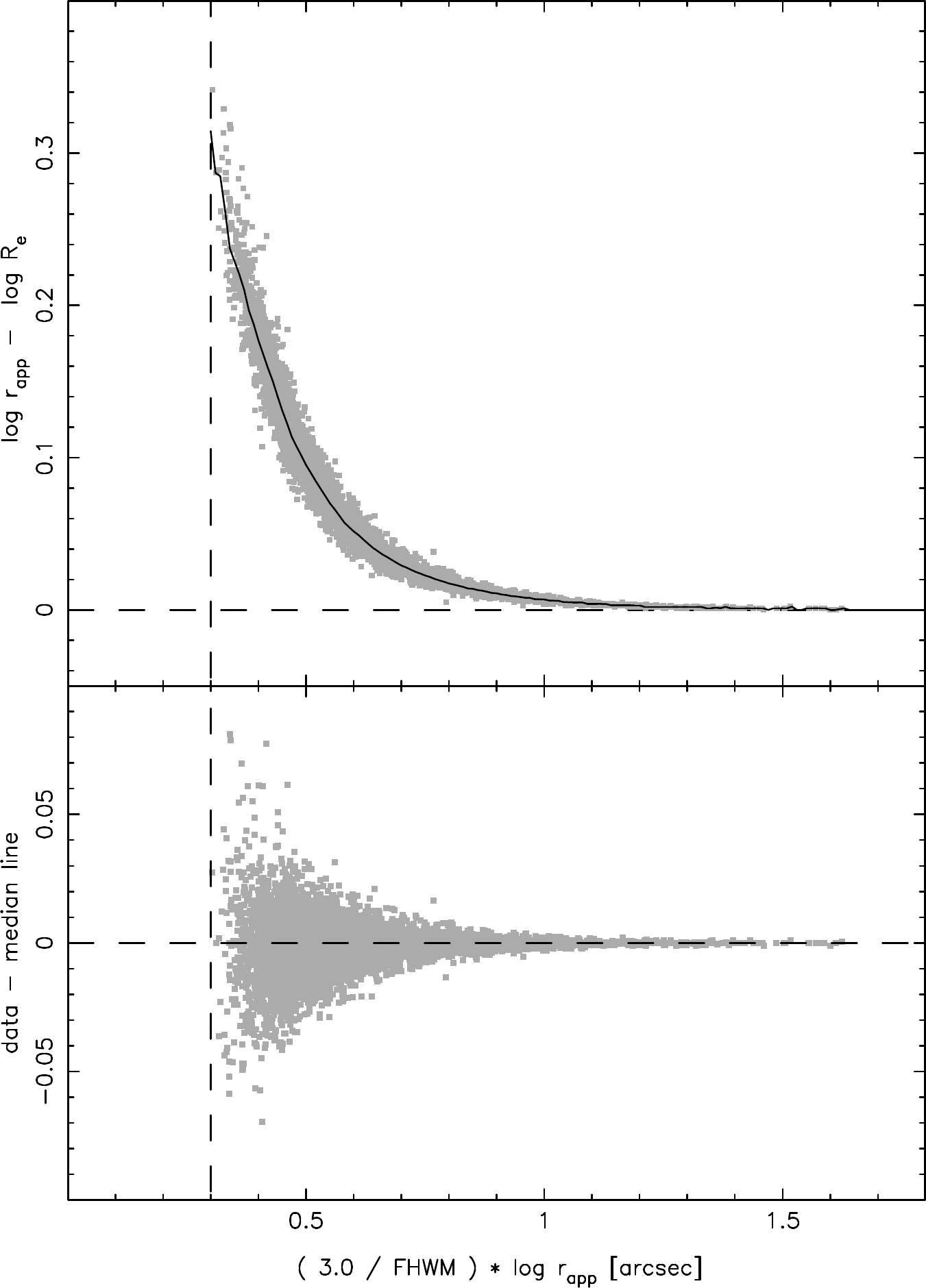} 
\caption{$J$-band PSF correction as determined by GALFIT modelling. The
  upper panel displays the measured PSF correction as a function of
  $r_{\rm app}$ normalised by the PSF FWHM on the parent 2MASS data
  tile. The vertical dashed line is where $r_{\rm app}=2$\,arcsec.
  The black line in the upper panel follows the median values.
  For values of (3.0\,/\,FWHM)\,log\,$r_{app}$ of 0.3, 0.4, ..., 1.4 the
  log\,$r_{app}$\,--\,log\,$R_{e}$\ corrections are 0.3143, 0.1768, 0.0950,
  0.0515, 0.0291, 0.0173, 0.0108, 0.0068, 0.0040, 0.0028, 0.0020, 0.0010
  respectively.
  The lower panel displays the residual between a median trend and
  the data.
}
\label{fig:r_cor}
\end{figure}

The 2MASS point-spread function (PSF) has a relatively large FWHM
($\sim$3.2\,arcsec) and required the development of a procedure to
derive PSF-corrected photometric parameters. For each target galaxy we
analysed the pixel data provided by the 2MASS Extended Source Image
Server as follows. The 2MASS image data for the $J$, $H$ and $K$ bands
were analysed independently. We adopted the total apparent magnitude
($m_T$) reported by 2MASS from the `fit extrapolation' method (i.e.\
{\tt j\_m\_ext}, {\tt h\_m\_ext}, {\tt k\_m\_ext}; see
\citealt{jarrett00}). First we determined the circular apparent
effective radius ($r_{\rm app}$) of the target galaxy on the 2MASS image
by finding, via interpolation, the radius that contained half the total
flux. A model two-dimensional Gaussian PSF image was derived from stars
on the parent 2MASS data tile. GALFIT \citep{peng10} was used with the
galaxy image and model PSF image as inputs to find the best-fit
two-dimensional S\'{e}rsic model. The half-light radius was determined
for the S\'{e}rsic model before and after convolution with the PSF
($r_{\rm model}$ and $r_{\rm smodel}$ respectively) to find the PSF
correction ($\Delta r$) given by
\begin{equation}
 \Delta\,r = r_{\rm smodel} - r_{\rm model} ~.
\end{equation}
The PSF-corrected effective radius ($R_e$) was then derived via
\begin{equation}
R_e = r_{\rm app} - \Delta r ~.
\end{equation}
Hence the effective radius is determined as the empirical half-light
radius corrected for the effects of the PSF using the S\'{e}rsic model
fit. Note that the GALFIT model fit is only used to determine the PSF
correction. The size and variation of $\Delta r$ as a function of
$r_{\rm app}$ for the $J$-band is given in Figure~\ref{fig:r_cor}.

For $J$-band apparent effective radii of 10, 8, 6, 4, 3 and 2\,arcsec,
the average $\Delta r$ and its dispersion are (0.15, 0.04), (0.20,
0.04), (0.27, 0.05), (0.51, 0.12), (0.88, 0.21) and (1.87, 1.10)
respectively. As the 2MASS PSF is nearly Gaussian, these corrections are
very similar to those derived from smoothing a S\'{e}rsic model with
the appropriate 2D Gaussian. For galaxies with apparent effective radii
of 2\,arcsec, the PSF correction is of similar size and so the relative
uncertainty of the correction is near unity. Hence this is the 
practical limit where galaxy sizes can be
measured reliably on the 2MASS image tiles.
Only 2\% of the
6dFGSv FP sample have apparent effective radii less than 2.4\,arcsec,
where the average correction is 1.4\,arcsec 
and hence the 2MASS image data is suitable for our study.

The derived PSF-corrected $JHK$ effective radii for 11\,102 galaxies are
presented in Table~\ref{tab:photo_data}. For 213 galaxies listed in
Table~\ref{tab:sigma_data}, i.e.\ 1.9\% of the sample, derivation of
reliable photometric parameters was not possible due to either
deblending/masking issues or because $r_{\rm app}$ was less than
2\,arcsec.

\begin{figure}
\centering
\includegraphics[width=0.4\textwidth]{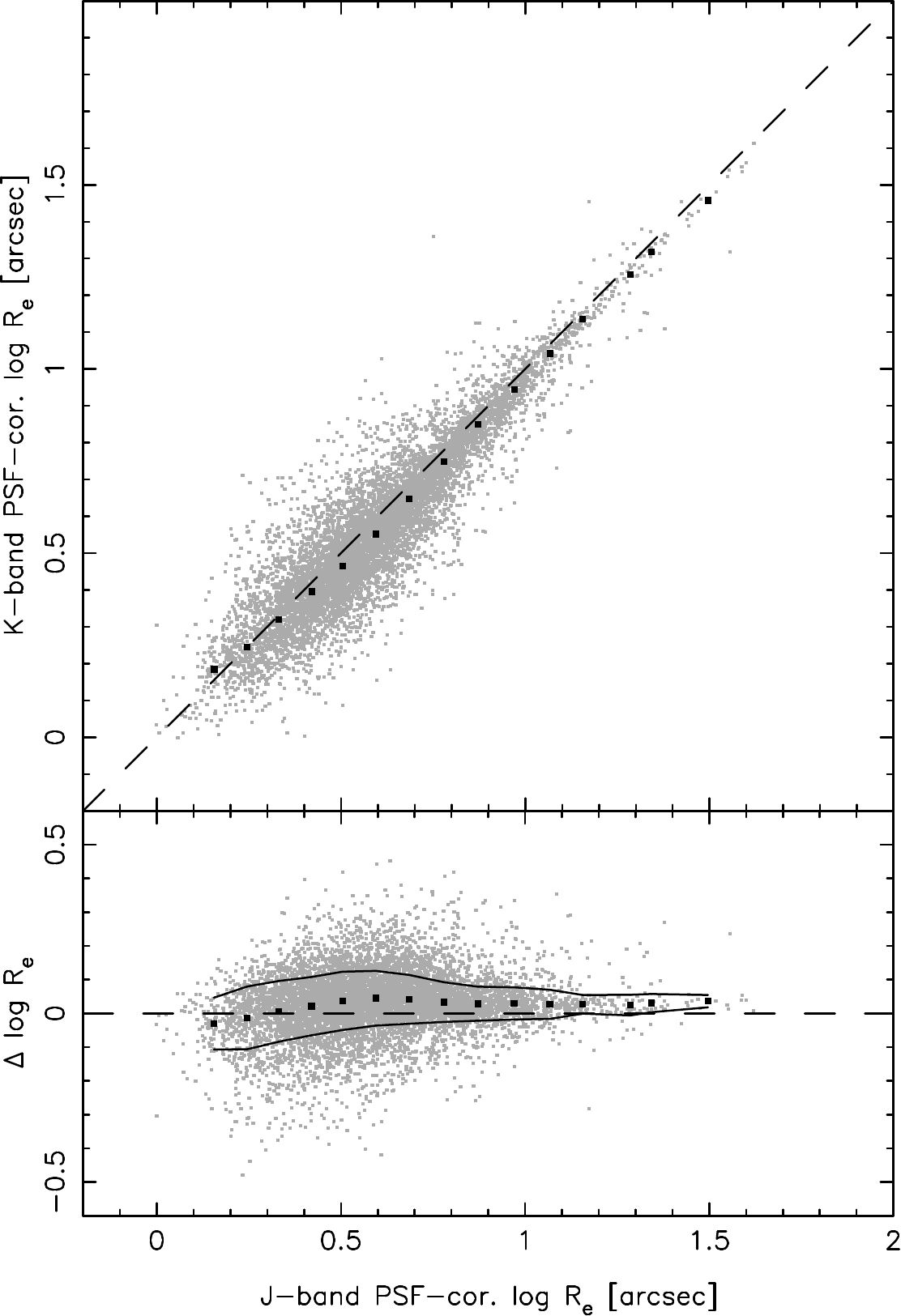} 
\caption{Comparison of the PSF-corrected $J$- and $K$-band measured
  $R_e$. The rms scatter of the difference is 0.078\,dex.
 The squares are the median values at evenly log\,$R_e$ steps. In the lower
 panel the lines show the locations of $\pm$ one standard deviations
 in the distribution.}
\label{fig:r_comp}
\end{figure}

The comparison of the PSF-corrected effective radii derived
independently from the $J$- and $K$-band 2MASS image data is presented
in Figure~\ref{fig:r_comp}. There is very good agreement between the two
sets of measurements. On average the measured $J$-band radii are
$\sim$7\% larger than those in the $K$-band. If the errors are equally
distributed between the two datasets, the average error on each
measurement is 12\%.

For each galaxy, the PSF-corrected effective surface brightness
$\langle \mu_e \rangle'$ is computed as
\begin{equation}
\langle \mu_e \rangle' = m_T + 2.5\log(2 \pi R_e^2)
\label{eqn:sb}
\end{equation}
where $m_T$ is the 2MASS total magnitude (e.g.\ {\tt j\_m\_ext}) and
$R_e$ is the PSF-corrected effective radius. In Section~5.2 we describe
corrections applied to $\langle \mu_e \rangle'$ for cosmological surface
brightness dimming, the k-correction, and galactic extinction in order
to derive the fully-corrected effective surface brightness $\langle
\mu_e \rangle$.

\begin{table*}
  \caption{6dFGSv 2MASS Photometric Parameters. 
    The columns are as follows:
    (1) 2MASS Extended Source Catalog name;
    (2) $J$-band total extended magnitude ({\tt j\_m\_ext});
    (3) $J$-band PSF FWHM in arcsec;
    (4) $J$-band $\Delta r$ correction applied in arcsec;
    (5) $\log$ $J$-band PSF-corrected circular effective radius in
        arcsec; the observed mean effective surface brightness can be
        derived via equation~\ref{eqn:sb}; 
    (6) $J$-band S\'{e}rsic index from GALFIT;
    (7) $H$-band total extended magnitude ({\tt h\_m\_ext});
    (8) $H$-band PSF FWHM in arcsec;
    (9) $H$-band $\Delta r$ correction applied in arcsec;
    (10) $\log$ $H$-band PSF-corrected circular effective radius in
         arcsec; 
    (11) $K$-band total extended magnitude ({\tt k\_m\_ext});
    (12) $K$-band PSF FWHM in arcsec;
    (13) $K$-band $\Delta r$ correction applied in arcsec;
    (14) $\log$ $K$-band PSF-corrected circular effective radius in
         arcsec. 
    The apparent magnitudes (j\_m\_ext, h\_m\_ext and k\_m\_ext) are 
   from the 2MASS Extended Source Catalog and are not corrected for galactic extinction.
    The full version of this table is provided in the online Supporting Information.
  }
\label{tab:photo_data}
{\scriptsize
\begin{tabular}{c ccccc cccc cccc}
\hline 
\hline

\multicolumn{1}{c}{  } & 
\multicolumn{5}{c}{ {\small {$J$-band}} } &
\multicolumn{4}{c}{ {\small {$H$-band}} } &
\multicolumn{4}{c}{ {\small {$K$-band}} } \\

\multicolumn{1}{c}{2MASS name}& 
\multicolumn{1}{c}{{\tt j\_m\_ext}}&
\multicolumn{1}{c}{FWHM}&
\multicolumn{1}{c}{${\Delta}r$ }&
\multicolumn{1}{c}{$\log{R_e}$ }&
\multicolumn{1}{c}{S\'{e}rsic $n$ }&
\multicolumn{1}{c}{{\tt h\_m\_ext} }&
\multicolumn{1}{c}{FWHM }&
\multicolumn{1}{c}{${\Delta}r$ }&
\multicolumn{1}{c}{$\log{R_e}$ }&
\multicolumn{1}{c}{{\tt k\_m\_ext} }&
\multicolumn{1}{c}{FWHM }&
\multicolumn{1}{c}{${\Delta}r$}&
\multicolumn{1}{c}{$\log{R_e}$}\\

\multicolumn{1}{c}{(1)} &  
\multicolumn{1}{c}{(2)} & 
\multicolumn{1}{c}{(3)} &
\multicolumn{1}{c}{(4)} &
\multicolumn{1}{c}{(5)} & 
\multicolumn{1}{c}{(6)} &
\multicolumn{1}{c}{(7)} & 
\multicolumn{1}{c}{(8)} &  
\multicolumn{1}{c}{(9)} & 
\multicolumn{1}{c}{(10)} &
\multicolumn{1}{c}{(11)} &
\multicolumn{1}{c}{(12)} & 
\multicolumn{1}{c}{(13)} & 
\multicolumn{1}{c}{(14)} \\

\hline
2MASXJ00001440-7652248 & 13.177 &  2.97 & 0.36 & 0.605 & 5.43 & 12.490 & 2.97 & 0.51 & 0.503 & 12.206  & 2.98 & 0.64 & 0.401\\
2MASXJ00002213-0137463 & 12.625 &  3.08 & 0.85 & 0.312 & 4.52 & 11.896 & 3.06 & 0.95 & 0.278 & 11.741  & 3.13 & 0.95 & 0.201\\
2MASXJ00002348-0656103 & 12.595 &  2.90 & 0.33 & 0.638 & 6.00 & 11.857 & 2.88 & 0.30 & 0.665 & 11.869  & 2.94 & 0.63 & 0.428\\
2MASXJ00002509-2602401 & 12.495 &  2.98 & 0.38 & 0.538 & 3.59 & 11.781 & 2.95 & 0.42 & 0.505 & 11.553  & 2.89 & 0.54 & 0.477\\
2MASXJ00003564-0145472 & 12.242 &  3.10 & 0.23 & 0.800 & 3.24 & 11.605 & 3.03 & 0.24 & 0.770 & 11.281  & 3.09 & 0.25 & 0.804\\
2MASXJ00003574-4034323 & 13.443 &  3.03 & 0.45 & 0.493 & 2.22 & 12.849 & 2.92 & 0.57 & 0.426 & 12.442  & 2.99 & 0.49 & 0.436\\
2MASXJ00004283-7217148 & 12.942 &  2.94 & 0.40 & 0.544 & 3.89 & 12.187 & 2.97 & 0.40 & 0.559 & 11.898  & 3.08 & 0.46 & 0.600\\
2MASXJ00004596-8158024 & 12.958 &  2.88 & 0.59 & 0.446 & 3.43 & 12.276 & 2.89 & 0.48 & 0.381 & 11.809  & 2.95 & 0.60 & 0.407\\
2MASXJ00004816-5511192 & 13.018 &  3.09 & 0.34 & 0.660 & 0.82 & 12.218 & 3.02 & 0.35 & 0.690 & 11.869  & 3.01 & 0.28 & 0.739\\
2MASXJ00005234-3550370 & 12.127 &  3.06 & 0.24 & 0.832 & 5.38 & 11.490 & 2.85 & 0.22 & 0.770 & 11.134  & 2.91 & 0.25 & 0.821\\
2MASXJ00005317-3559104 & 12.445 &  3.02 & 0.33 & 0.671 & 4.14 & 11.620 & 2.94 & 0.25 & 0.743 & 11.488  & 2.98 & 0.37 & 0.651\\
\hline

\end{tabular}
}
\end{table*}
\subsection{Internal $X_{FP}$ comparisons}
\begin{figure} 
\centering
\includegraphics[width=0.4\textwidth]{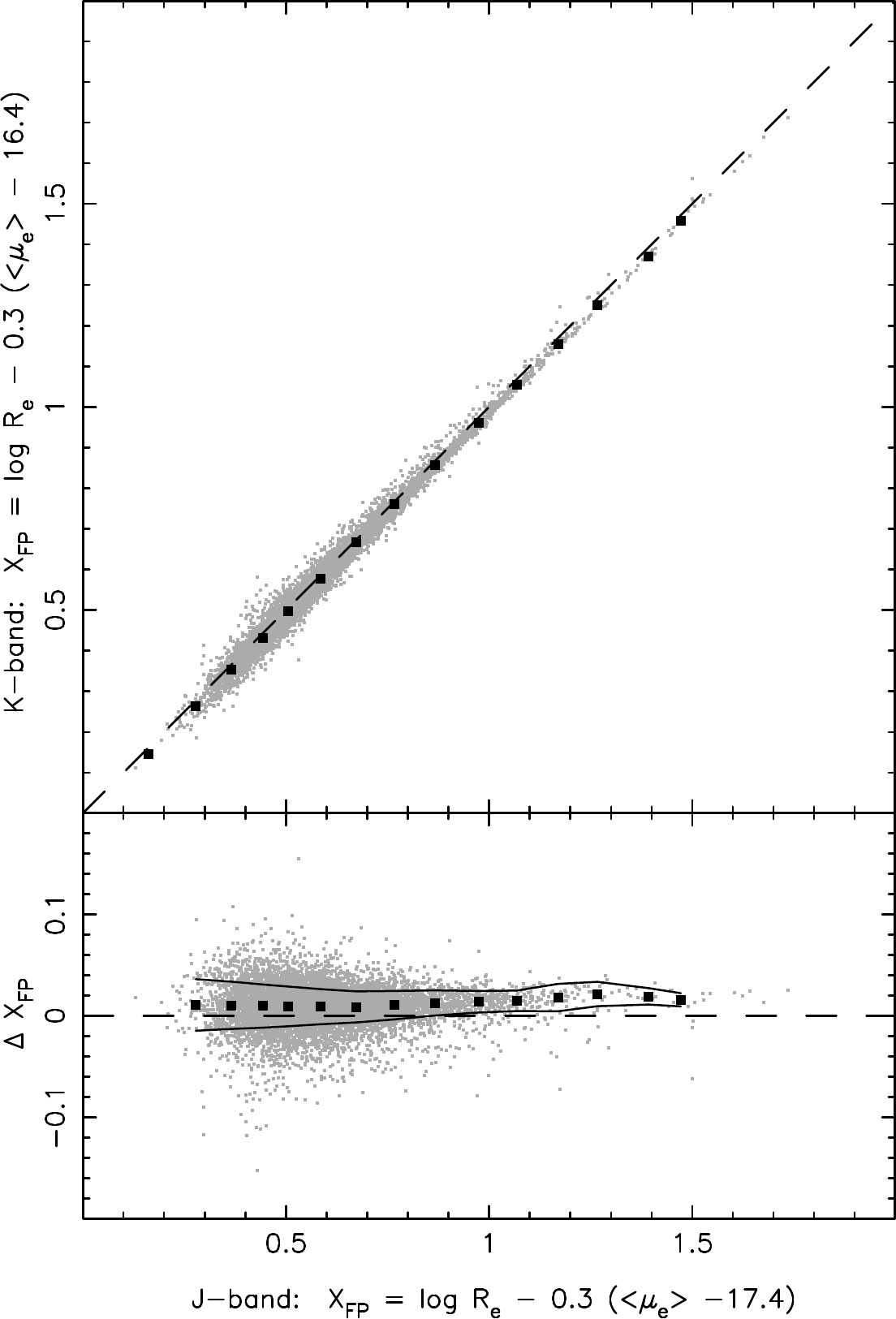} 
\caption{Comparison of $X_{FP}$ ($\equiv \log{R_e} - 0.3 \langle \mu_e
  \rangle$) for the $J$- and $K$-band measurements. 
  The average effective surface brightness for the sample in
  each band has been subtracted from the $\langle \mu_e \rangle$ values. 
  For this comparison we include corrections to
  the effective surface brightnesses for galactic extinction, the
  k-correction and cosmological surface brightness dimming (see Section 5.2).
 The squares are the median values at uniform $X_{FP}$ steps. 
In the lower
 panel the lines show the locations of $\pm$ one standard deviations
 in the distribution. The rms scatter of the difference is 0.018\,dex.}
\label{fig:xfp_comp}
\end{figure}

As is well-known the measurements of $R_e$ and $\langle \mu_e \rangle$ are highly correlated; if $R_e$ is overestimated then a fainter $\langle \mu_e
\rangle$ is derived, and conversely  {\citep[e.g.][]{burstein87,lucey91}}.
However the linear combination of
these parameters that appears in the FP, which is approximately
$X_{FP} \equiv R_e - 0.3 \langle \mu_e \rangle $, is well determined
\citep[e.g.][]{jorgensen95a}. 
In  Figure~\ref{fig:xfp_comp} we compare
our $X_{FP}$ measurements for the $J$- and $K$-bands.
To aid the comparison for each band the average effective surface 
brightness has been subtracted from the $\langle \mu_e \rangle$ values
which ensures that the average $\Delta\,X_{FP}$ is near zero.
The low intrinsic scatter between these two independent measurements 
implies excellent internal precision for each dataset
(regardless of any intrinsic colour-size relation).
The implied average measurement error of $X_{FP}$ , if equally divided
between the two datasets, is 0.013\,dex (3\%).

\begin{figure} 
\centering
\includegraphics[width=0.4\textwidth]{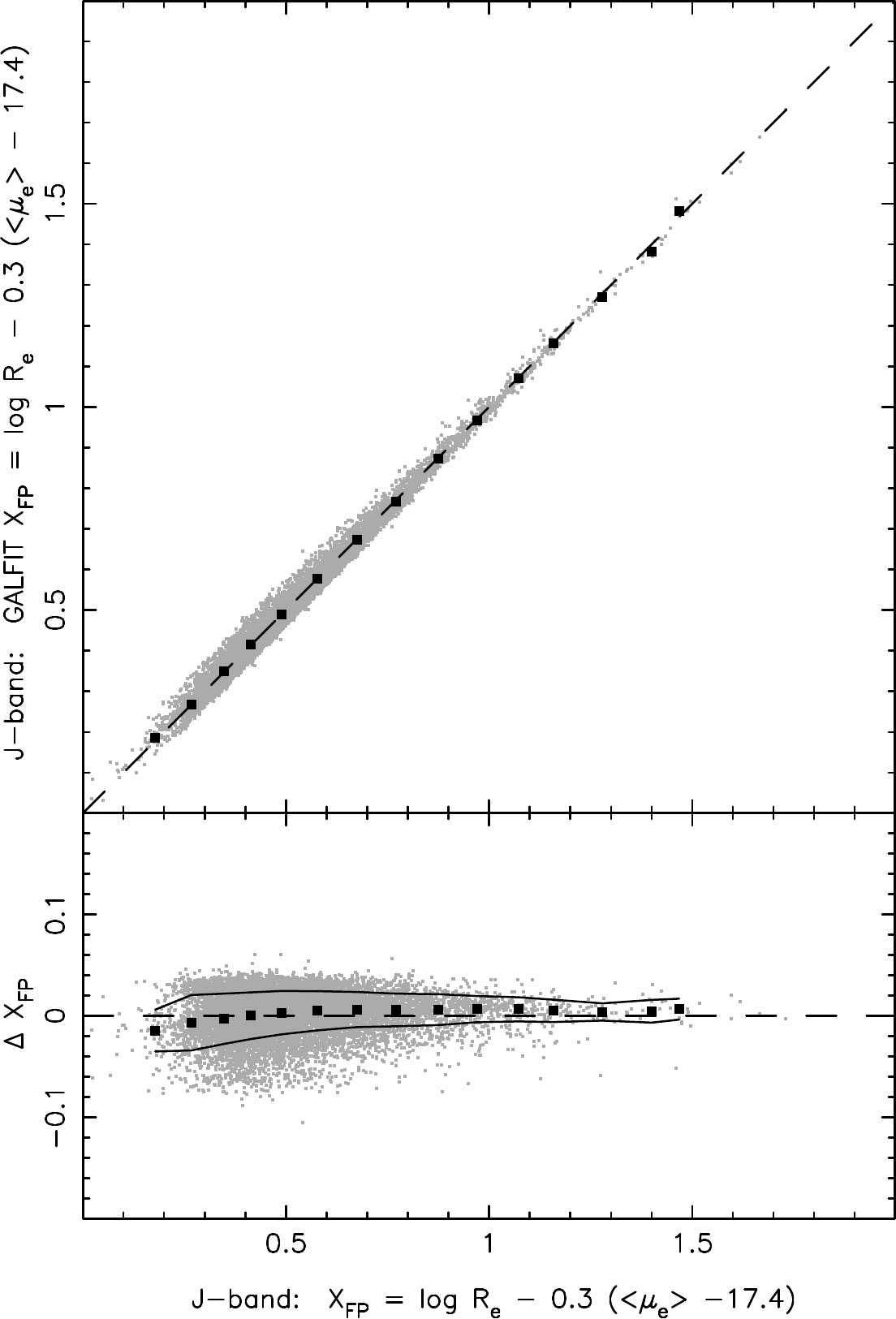} 
\caption{
Comparison of $X_{FP}$ 
  derived from our technique and directly with GALFIT. To aid the
  comparison, the average effective surface brightness for the sample in
  each band has been subtracted from $\langle \mu_e \rangle$. 
 The squares are the median values at uniform $X_{FP}$ steps. 
 In the lower
  panel the lines show the locations of $\pm$ one standard deviations
  in the distribution. The rms scatter of the difference is 0.021\,dex.}
\label{fig:xfp_galfit_comp}
\end{figure}

Our analysis used the total apparent magnitude reported by
2MASS, e.g. {\tt j\_m\_ext}, and PSF-corrected apparent
effective radius derived using GALFIT.
An alternative approach would have been to directly use the total 
S\'{e}rsic magnitude and effective radii returned by GALFIT.
The $X_{FP}$ parameters from these two techniques are in excellent
agreement (see Figure~\ref{fig:xfp_galfit_comp}) with an average
difference of 0.001 dex and the scatter in the difference of 0.021 dex.
Hence for the relatively low S/N 2MASS image data these two 
techniques are equivalent. In our analysis we prefer
our approach as this does not force a particular parametric
photometric form on the galaxy light distribution.

\subsection{External $X_{FP}$ comparisons}
\begin{figure} 
\centering
\includegraphics[width=0.4\textwidth]{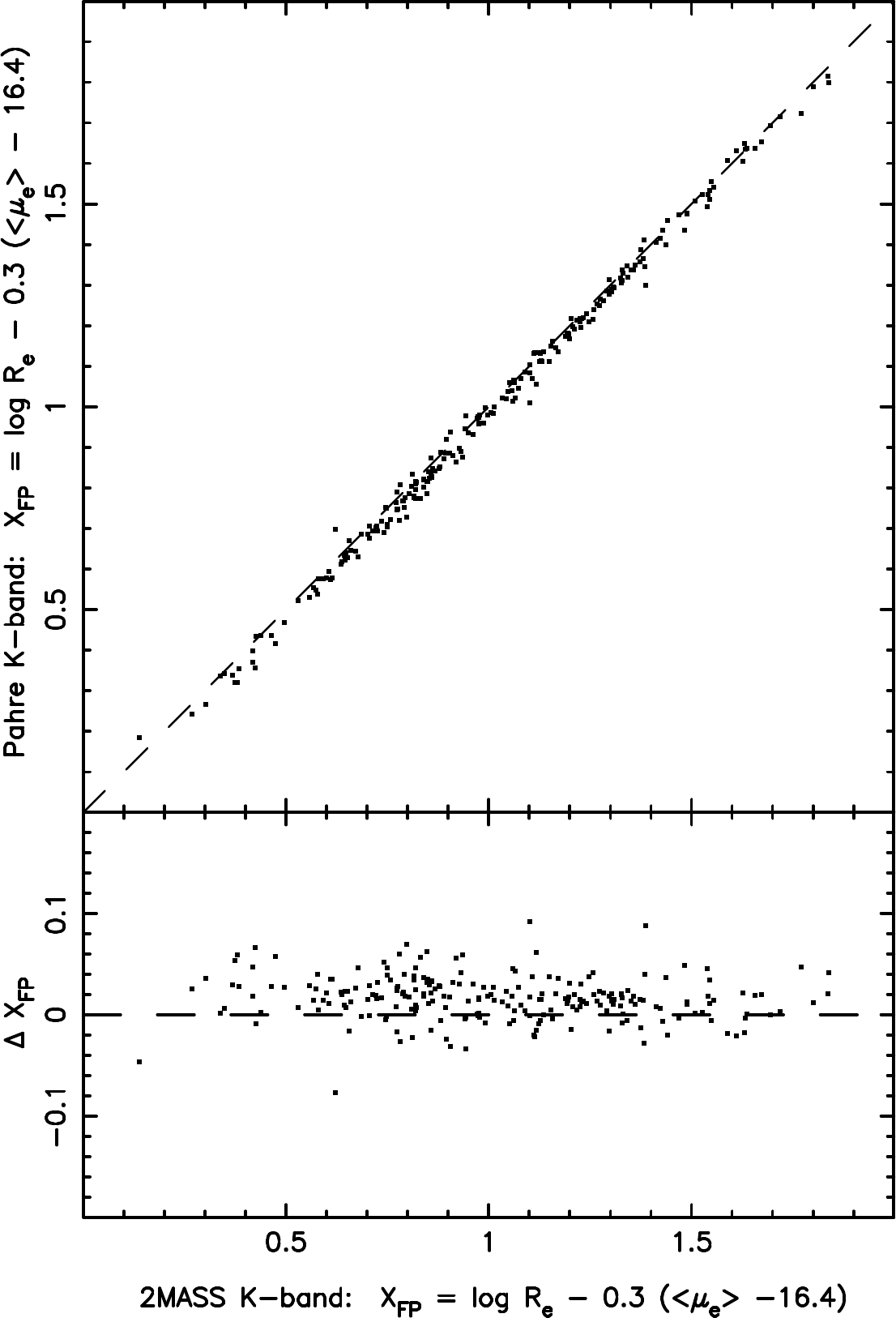}
\caption{
Comparison of the $K$-band $X_{FP}$
for our 2MASS-based measurements and the Pahre measurements.
The observed median offset (2MASS-Pahre) is 0.017 dex (equivalent
to a magnitude offset of 0.05) and the rms scatter is 0.020 dex.
}
\label{fig:xfp_pahre_comp}
\end{figure}

\cite{pahre99} measured K-band FP photometric parameter for 341 nearby
early-type galaxies primarily in clusters.
We have analysed the 2MASS $K$-band image data following the procedure outlined above 
and measured $R_e$ and $\langle \mu_e \rangle$ for the Pahre sample.
2MASS-based measurements could be derived for 229 galaxies; 
the missing galaxies are either too faint, unresolved 
at the 2MASS resolution or strongly contaminated by an
adjacent star. The comparison of the $X_{FP}$ measurements is presented in
Figure~\ref{fig:xfp_pahre_comp}. There is excellent agreement between
the two datasets with a systematic offset of 0.017 dex.
The implied average measurement error, if equally divided between the two dataset,
is 0.014 dex (3\%).

\begin{figure} 
\centering
\includegraphics[width=0.4\textwidth]{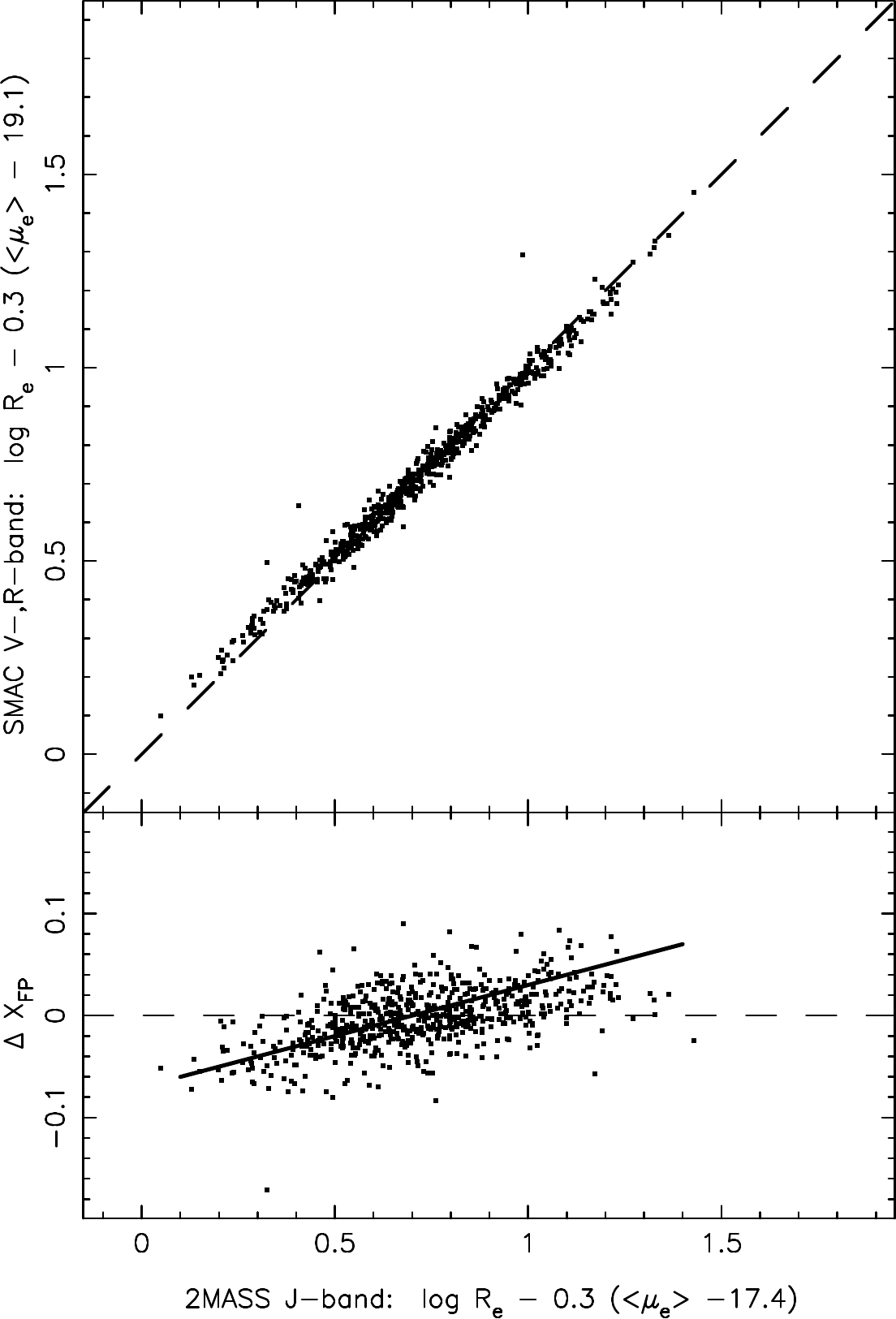}
\caption{
  Comparison of the $X_{FP}$ for the 2MASS $J$-band and SMAC $V$-, $R$-band
  measurements. To aid the comparison the average effective surface
  brightness for the sample in each band has been subtracted from
  $\langle \mu_e \rangle$. For this comparison we include corrections to
  the effective surface brightnesses for galactic extinction, the
  k-correction and cosmological surface brightness dimming. The rms
  scatter of the difference is 0.027\,dex. 
  The residual trend (lower panel) is a direct result of the 
  colour-magnitude relation and is detected at high significance.
  The slope of the $J\,-(V,R)$ colour-magnitude for the SMAC sample
  is $\sim$0.1 and a line of slope 0.1 is shown in the lower panel,
  i.e. this is the expected systematic trend that would result from
  the observed colour-magnitude relation.
}
\label{fig:xfp_smac_comp}
\end{figure}

The SMAC survey (\citealt{hudson04} and references therein) presented
FP photometric parameters derived from either
$R$- or $V$-band data an all-sky sample of 699
early-type galaxies in 56 clusters. For all galaxies in the SMAC
standardized FP catalogue (\citealt{hudson01}, Table 7) we have analysed
the 2MASS $J$-band image data following the procedure outlined above and
measured $R_e$ and $\langle \mu_e \rangle$. 2MASS-based measurements
could be derived for 668 galaxies. On average $R_e^{J}/R_e^{V,R} = 0.70$. 
The comparison of
the $X_{FP}$ measurements is presented in
Figure~\ref{fig:xfp_smac_comp}. 
In this comparison the average effective surface brightness in each 
band has been subtracted from the $\langle \mu_e \rangle$ values. 
The low intrinsic scatter between these two fully independent sets of 
measurements implies excellent internal precision for each dataset. 
The implied average measurement error of $X_{FP}$ for 
each band is 0.019\,dex (4\%).
The residual trend in the lower panel is a direct result of 
the colour-magnitude relation; at fixed $\log{R_{e}(J)/R_{e}(R)}$,
  $ \Delta X_{FP} \propto $ 
  $\langle \mu_{e}(J) \rangle - \langle \mu_{e}(R) \rangle$ 
  $\propto J\,-R$ colour and  $X_{FP}\,\propto\,J$ magnitude.

\begin{figure} \centering
\includegraphics[width=0.4\textwidth]{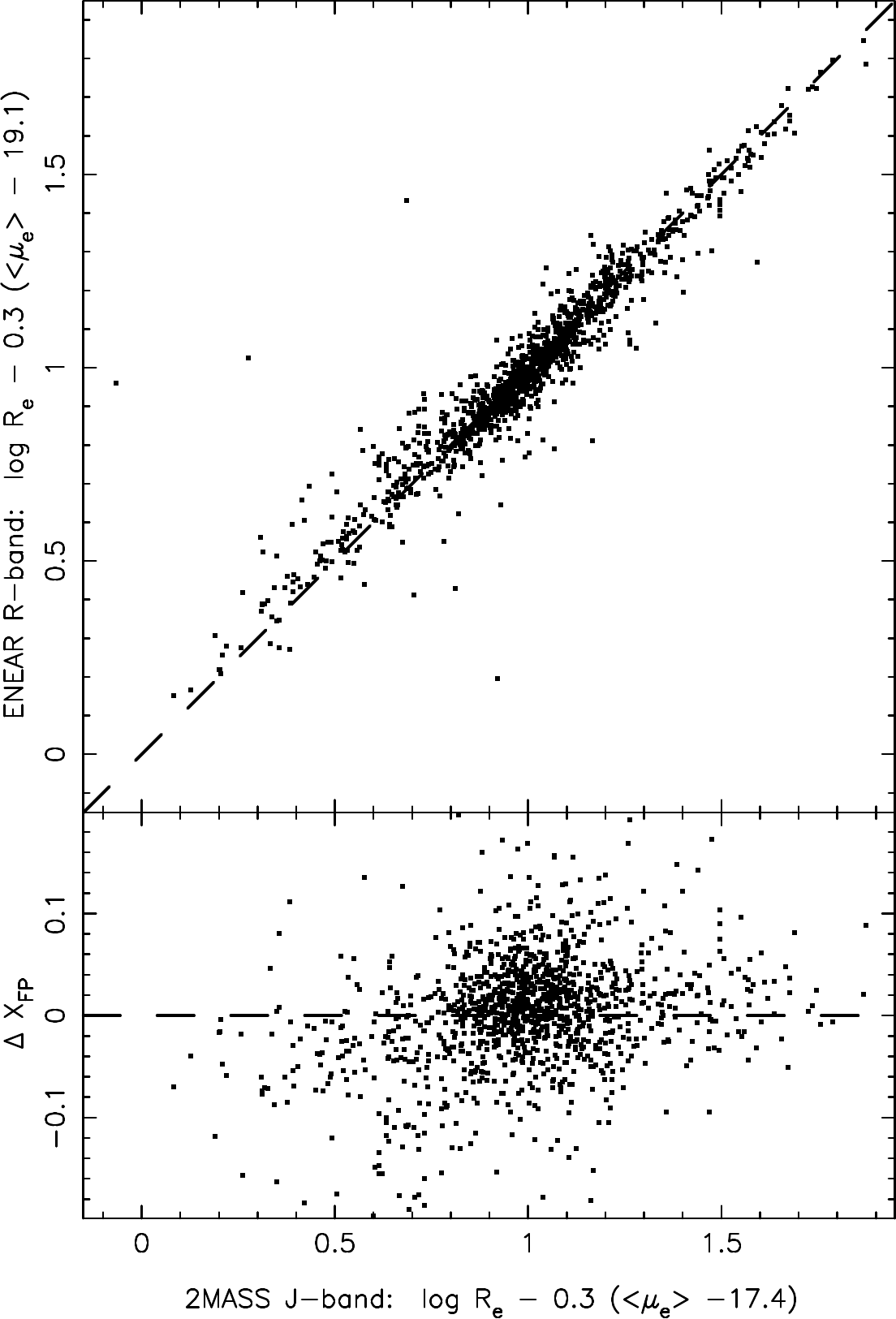}
\caption{Comparison of the $X_{FP}$ for the 2MASS $J$-band and ENEAR $R$-band
  measurements. To aid the comparison the average effective surface
  brightness for the sample in each band has been subtracted from
  $\langle \mu_e \rangle$. For this comparison we include corrections to
  the effective surface brightness for galactic extinction, the
  k-correction and cosmological surface brightness dimming. The rms
  scatter of the difference is 0.047\,dex.
}
\label{fig:xfp_enear_comp}
\end{figure}

The ENEAR survey (\citealt{dacosta00}) presented $R$-band FP photometric
parameters for an all-sky sample of 1332 early-type galaxies. For all
galaxies listed in the ENEAR photometric catalogue \citep{alonso03} we
have analysed the 2MASS $J$-band image data following the procedure
outlined above and measured $R_e$ and $\langle \mu_e \rangle$.
2MASS-based measurements could be derived for 1270 galaxies. The
comparison of the $X_{FP}$ measurements is presented in
Figure~\ref{fig:xfp_enear_comp}. There is good agreement between the two
datasets, with an implied average measurement error of $X_{FP}$ for each
band of 0.033 dex (7.5\%).

In \citet{magoulas12} (see Section~3.3) we describe in detail how the
uncertainties in $R_e$ and $\langle \mu_e \rangle$ (and their correlated
nature) are taken into account when we apply our maximum likelihood
method to model the 3D FP.

\section{Visual Classification}
\newlength{\plotwidth}
\newlength{\fullwidth}
\setlength{\plotwidth}{\columnwidth}            
\setlength{\fullwidth}{\textwidth}              
\setlength{\tabcolsep}{1ex}
\newcommand{\eg}{\mbox{\it e.g.}}
\newcommand{\tdfdr}{\mbox{\tt 2dfdr}}
\newcommand{\h}{\mbox{$^{\rm h}$}}
\newcommand{\m}{\mbox{$^{\rm m}$}}
\newcommand{\s}{\mbox{$^{\rm s}$}}
\newcommand{\persqdeg}{\mbox{\,deg$^{-2}$}}
\newcommand{\permag}{\mbox{\,mag$^{-1}$}}
\newcommand{\magdeg}{\mbox{\,\permag\persqdeg}}
\newcommand{\sqdeg}{\mbox{\,deg$^{2}$}}
\newcommand{\perpix}{\mbox{\,pixel$^{-1}$}}
\newcommand{\Mpc}{\mbox{$\,h^{-1}\,{\rm Mpc}$}}
\newcommand{\Gpc}{\mbox{$\,h^{-1}\,{\rm Gpc}$}}
\newcommand{\parsec}{\mbox{$\,h^{-1}\,{\rm pc}$}}
\newcommand{\kiloparsec}{\mbox{$\,h^{-1}\,{\rm kpc}$}}
\newcommand{\cubicMpc}{\mbox{$\,h^{-3}\,{\rm Mpc}^3$}}
\newcommand{\cubicGpc}{\mbox{$\,h^{-3}\,{\rm Gpc}^3$}}
\newcommand{\invMpc}{\mbox{$\,h\,{\rm Mpc}^{-1}$}}
\newcommand{\invcubicMpc}{\mbox{$\,h^3\,{\rm Mpc}^{-3}$}}
\newcommand{\invcubicGpc}{\mbox{$\,h^3\,{\rm Gpc}^{-3}$}}
\newcommand{\kmsMpc}{\mbox{\,km\,s$^{-1}$\,Mpc$^{-1}$}}
\newcommand{\CF}{\mbox{$\rm c_{\rm F}$}} 
\newcommand{\head}[1]{\multicolumn{1}{c}{#1}}
\newcommand{\n}{\phantom{0}}
\newcommand{\phpm}{\phantom{$\pm$}}
\newcommand{\notice}[1]{{\it [{#1}]}}
\newcommand{\xx}{\scriptsize \tt}

\newcommand{\kb}{\mbox{$K$}}
\newcommand{\hb}{\mbox{$H$}}
\newcommand{\jb}{\mbox{$J$}}
\newcommand{\bj}{\mbox{$b_{\rm\scriptscriptstyle J}$}}
\newcommand{\rf}{\mbox{$r_{\rm\scriptscriptstyle F}$}}

\newcommand{\dkb}{\mbox{$\Delta K$}}
\newcommand{\dhb}{\mbox{$\Delta H$}}
\newcommand{\djb}{\mbox{$\Delta J$}}
\newcommand{\dbj}{\mbox{$\Delta b_{\rm\scriptscriptstyle J}$}}
\newcommand{\drf}{\mbox{$\Delta r_{\rm\scriptscriptstyle F}$}}

\newcommand{\khjrb}{\mbox{$KHJr_{\rm\scriptscriptstyle F}b_{\rm\scriptscriptstyle J}$}}
\newcommand{\brjhk}{\mbox{$b_{\rm\scriptscriptstyle J}r_{\rm\scriptscriptstyle F}JHK$}}

\newcommand{\mlim}{\mbox{$m_{\rm lim}$}}
\newcommand{\ktot}{\mbox{$K_{\rm tot}$}}
\newcommand{\kiso}{\mbox{$K_{\rm iso}$}}
\newcommand{\mk}{\mbox{$M_K$}}
\newcommand{\mh}{\mbox{$M_H$}}
\newcommand{\mj}{\mbox{$M_J$}}
\newcommand{\mr}{\mbox{$M_r$}}
\newcommand{\mb}{\mbox{$M_b$}}
\newcommand{\bmk}{\mbox{$(b_{\rm\scriptscriptstyle J}-K)$}}
\newcommand{\MbmMk}{\mbox{$(M_b - M_K)$}}

\newcommand{\al}{\mbox{$\alpha$}}
\newcommand{\ms}{\mbox{$M^*$}}
\newcommand{\ps}{\mbox{$\phi^*$}}
\newcommand{\ls}{\mbox{$L^*$}}
\newcommand{\pc}{\mbox{$\phi_{\rm conv}$}}

\newcommand{\vmax}{\mbox{$1/V_{\rm max}$}}

\newcommand{\plotone}[1]
  {\centering \leavevmode \includegraphics[clip,width=\plotwidth]{file=#1}} 

\newcommand{\plotfull}[2]
  {\centering \leavevmode \includegraphics[clip,width=#2\fullwidth]{file=#1}}

\newcommand{\dummyfigure}
    {\fbox{\parbox{0.95\plotwidth}{\centering\it\vspace*{6.5cm}
     Figure inserted about here in text.\vspace*{6.5cm}}}} 

\newcommand{\dummytable}
    {\fbox{\parbox{0.95\plotwidth}{\centering\it\vspace*{0.5cm}
     Table inserted about here in text.\vspace*{0.5cm}}}} 

\subsection{Overview}

\begin{figure*}
\includegraphics[clip,width=0.8\fullwidth]{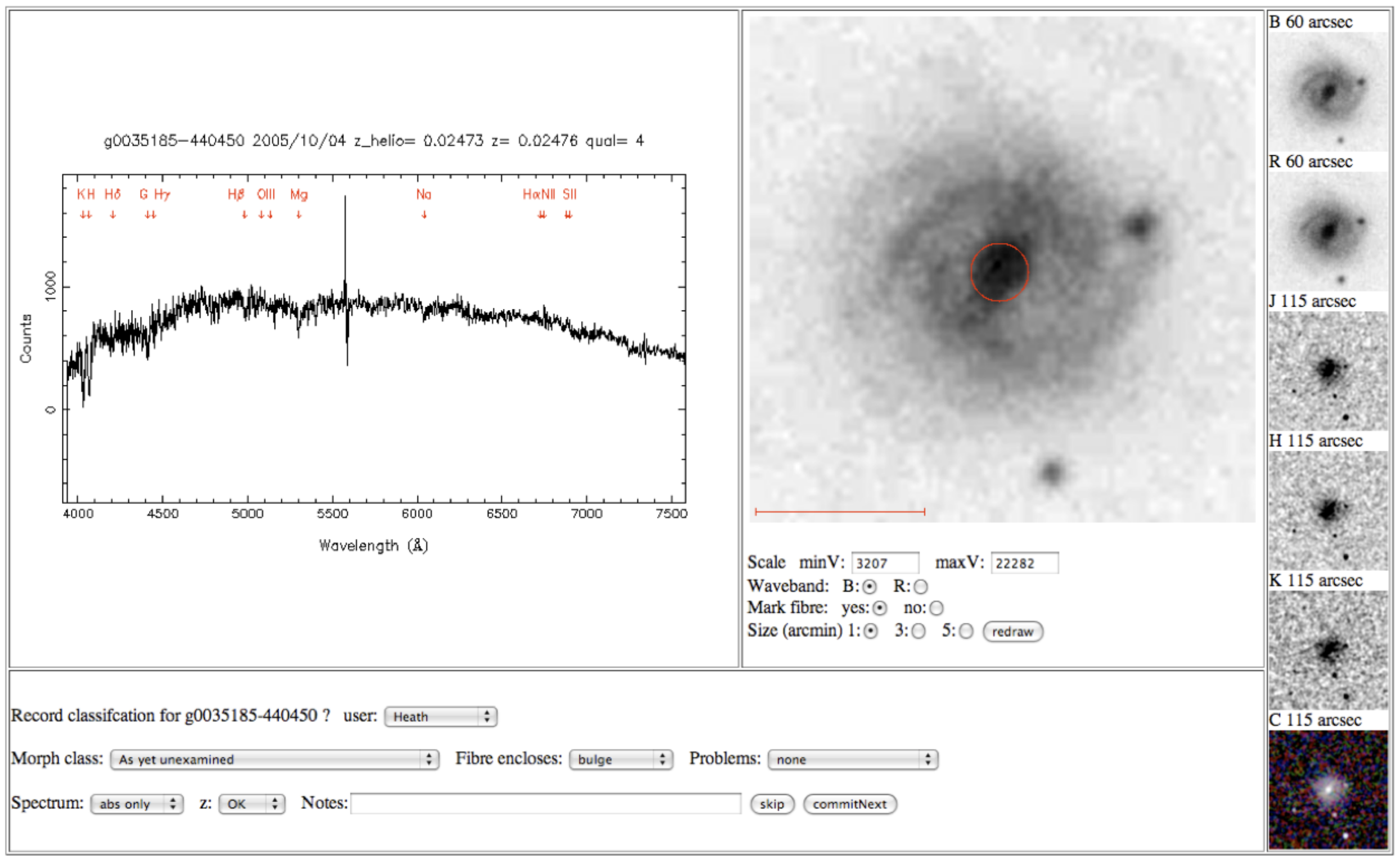} 
\caption{Purpose-built interface used to morphologically classify
  galaxies. The example shown is the nearby barred-spiral
  ESO~242$-$G\,017 ($z=0.02473$). This galaxy is an example of a
  late-type system with a bulge that fills the 6dF fibre aperture,
  giving an early-type spectrum with negligible line emission.}
\label{figmorph0}
\end{figure*}

All galaxies in the 6dFGSv sample were examined visually 
to verify that they were bulge-dominated systems, 
appropriate for the FP analysis. Moreover, we attempted to
roughly classify the types of galaxies in the sample
by Hubble Type, as well as to identify cases which were problematic
either in Type or in measurement.
Visual classification was
undertaken via a private purpose-built interface to the
publicly-accessible 6dFGS online database. The interface, constructed by
M.\,Read, recorded information input by our group in SQL format 
so they could be linked to the original database
entries for each galaxy. To this end, the interface showed scaled $\jhk$
images from the 2MASS extended Source Catalog \citep[XSC;][]{jarrett00xsc} 
and $\br$ images from SuperCOSMOS \citep{hambly01}.
These images were displayed alongside the redshifted-labelled spectrum
for the galaxy (example shown in Fig.~\ref{figmorph0}).

\begin{figure}
\includegraphics[clip,width=\plotwidth]{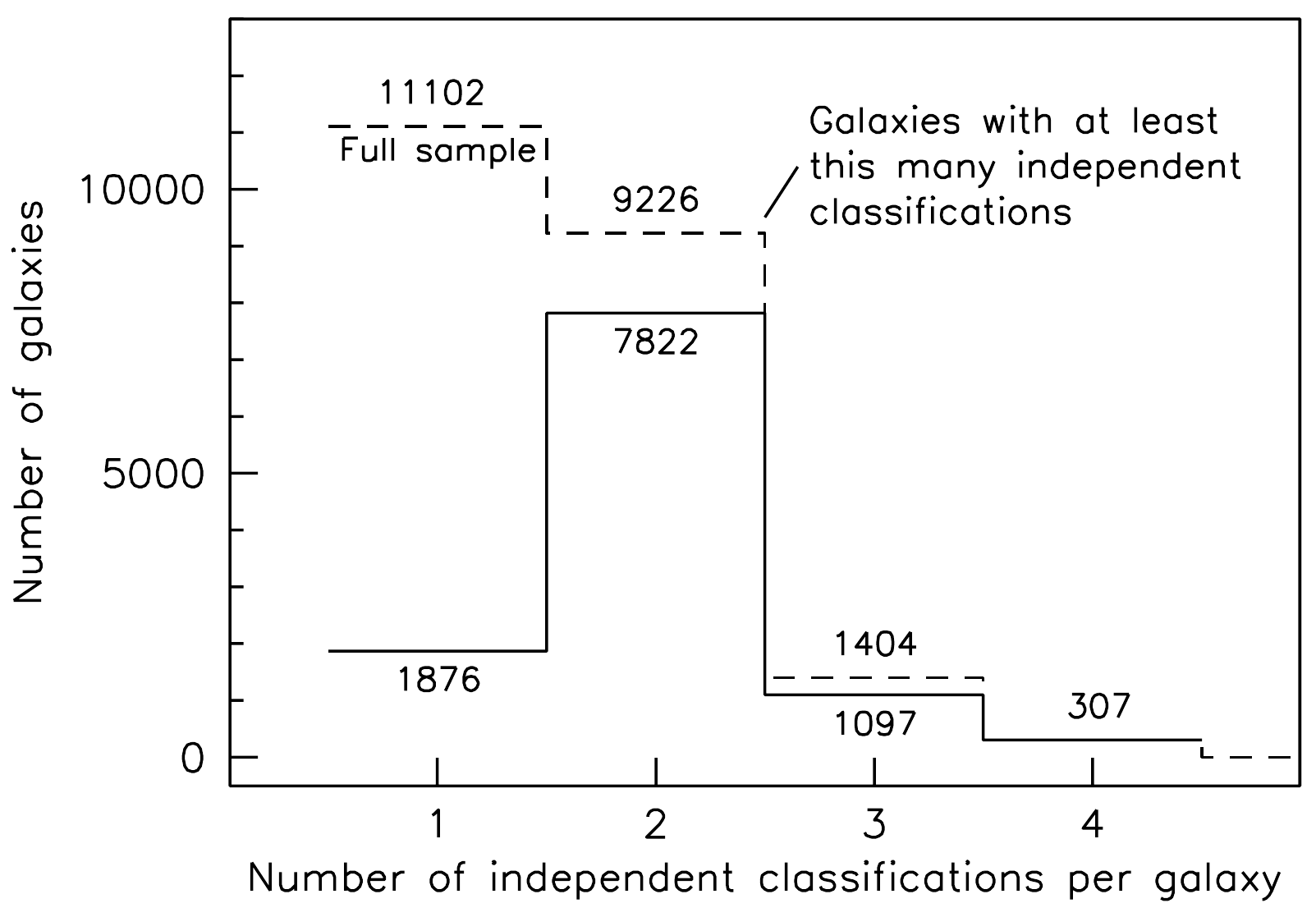} 
\caption{Differential ({\it solid}) and cumulative ({\it dashed})
  distributions of the number of galaxies with a given number of
  morphological classifications. By {\it classifications}, we mean the
  number of times an individual galaxy was independently viewed and
  assessed by one of our group. The cumulate runs from right to left and
  represents the number of sample galaxies having at least that number
  of independent classifications. 
}
\label{figmorph1}
\end{figure}

The classification work was undertaken by ten members of the 6dFGS project
team\footnote{ Beutler,
Cluver, Colless, Jarrett, Jones, Lucey, Magoulas, Mould, Parker, and
Springob} and was organised to maximise the number of galaxies viewed
and classified independently by at least two people.
Figure~\ref{figmorph1} shows the number of galaxies viewed a given
number of times by independent classifiers. It shows that there were
9\,226 galaxies (83\% of the full sample) classified two or more times.
Repeating the classifications using different people for most of
the sample means that the dispersion in the type assignments can be used
as a measure of self-consistency.

Our classification scheme was modelled on the one used by
\citet{dressler80}, who targetted galaxies over comparable distances
with similar resolution and sensitivity. Like all modern qualitative
schemes, ours was based on the \citet{de59} revision of Hubble's
original morphological sequence \citep{hubble26}. Although we did not
distinguish between barred and normal spirals, we did separate galaxies
into the major categories of elliptical (E), lenticular (S0), spiral
(Sp), and irregular (Irr) types, as well as intermediate cases
\citep[e.g. ][]{sandage61,jarrett00}.

Table~\ref{tab:morph} lists the classification divisions used in our
analysis. We assigned numerical values (called $m$-types) running from
$-4$ to $+6$ on a scale of our own invention. Galaxy $m$-types ($m$ for
morphology) differ from other numerical classification schemes (notably
$T$-types) in the way that galaxy orientation was encoded. The problem
is not so much orientation per se, but the way in which the inclination
affects different types disproportionally: disk galaxies are more likely
to be misclassified at high inclination angles because of their
flattened shape.

\begin{table}
\begin{center}
  \caption{Morphological classification of the 6dFGSv \label{tab:morph}} 
\begin{tabular}{lc}
\hline
Morphology  & $m$-type \\
\hline
& \\
Edge-on disk galaxy & $-4$ \\
with full-length dust lane & \\
& \\
Edge-on disk galaxy with & $-2$ \\
partial or no dust lane & \\
& \\
Elliptical & 0 \\
& \\
Transition case: & $+1$ \\
Elliptical -- S0 & \\
& \\   
S0 (inclined/face-on) & $+2$ \\
&  \\
Transition case: & $+3$ \\
S0 -- Spiral & \\
&  \\
Spiral (inclined/face-on) & $+4$ \\
&  \\
Transition case: & $+5$ \\
Spiral -- Amorphous & \\
& \\
Irregular or Amorphous & $+6$ \\
& \\
Unclassifiable & {\sl 999} \\
& \\
\hline
\end{tabular}\\
\end{center}
\end{table}

The established morphological $T$-type scale \citep{de76,simien86}
assigns values from $-6$ for ellipticals up to $+7$ for late-type
spirals (Sc/Sd) and beyond ($\geq 8$) for irregular galaxies. Negative
$T$-types refer exclusively to early types (ellipticals/lenticulars)
while late-types (spirals/irregulars) have positive values.
\citet{fukugita07} used their own variation on $T$-types with a scale
from 0 to 6 (with $-1$ reserved for unclassifiable objects).

A galaxy in our scheme with $m = 0$ is an ordinary elliptical.
Unambiguous, low-inclination-angle ($i \lesssim 60^\circ$) disk galaxies
were assigned $m=+2$ in the case of lenticulars (S0s) and $m=+4$ for
ordinary spirals. No attempt was made to subdivide spirals into early
and late types. Irregular galaxies were allocated $m=+6$. Face-on and
low-inclination S0s are distinguished from Es by the presence of a
bulge. For the purpose of deriving a clean FP, our ability to
distinguish spirals from S0s is paramount, much more so than separating
S0s and ellipticals.

The intermediate odd-numbered values ($m=+1,+3,+5$) were used to denote
the transition or intermediate cases between the key types. Transition galaxies are
those that are ambiguous in appearance, regardless of whether this
reflects their true nature. Because of this, our use of the word 
{\it transitional} is not meant to imply anything about how galaxies
evolve.

Most importantly, negative $m$-types were reserved for edge-on cases
(inclination angles $i \gtrsim 60^\circ$) where the galaxy has an
obvious disk but its nature as an S0 or spiral is ambiguous. Galaxies
with $m=-4$ were edge-on cases with an obvious dust lane, while $m=-2$
is given to those edge-on cases with no obvious dust lane. Although our
use of optical $\br$ imaging facilitated these distinctions, edge-on
dust lanes are not an unequivocal indicator of spiral structure. However
the $m$-type scheme incorporates this limitation by allowing edge-on
cases to be flagged.

The utility of the $m$-type scheme is that it makes provision for
inclination while simultaneously preserving classification information.
By defining $|m|$-type as the absolute value of the $m$-type, edge-on
disk galaxies (with $m = -2, -4$) are readily combined with all other
disk galaxies in the sample (values $m = +2, +4$) if one so chooses.
Alternatively, if we want to remove the edge-on cases entirely, we can
do that by simply restricting the sample to $m \geq 0$.

The advantages of this approach were twofold. First, it acknowledges the
unavoidable role played by inclination by incorporating it into the
classification scheme. Secondly, it affords a simple way of
differentiating between galaxies for which orientation has played a role
from those of the same morphological type on a surer footing. In this
way we deferred any decision about whether a particular classification
was incorporated into our final sample. This ensured that the
decision-making process imposed on classifiers was kept as simple as
possible, enabling a speedy and consistent classification of thousands
of galaxies. Problem cases were also tagged, ensuring a clean final
sample.  There were 389 galaxies (3.5\% of the sample) for which two or
more classifiers flagged a problem.

\begin{figure}
\includegraphics[width=0.45\textwidth]{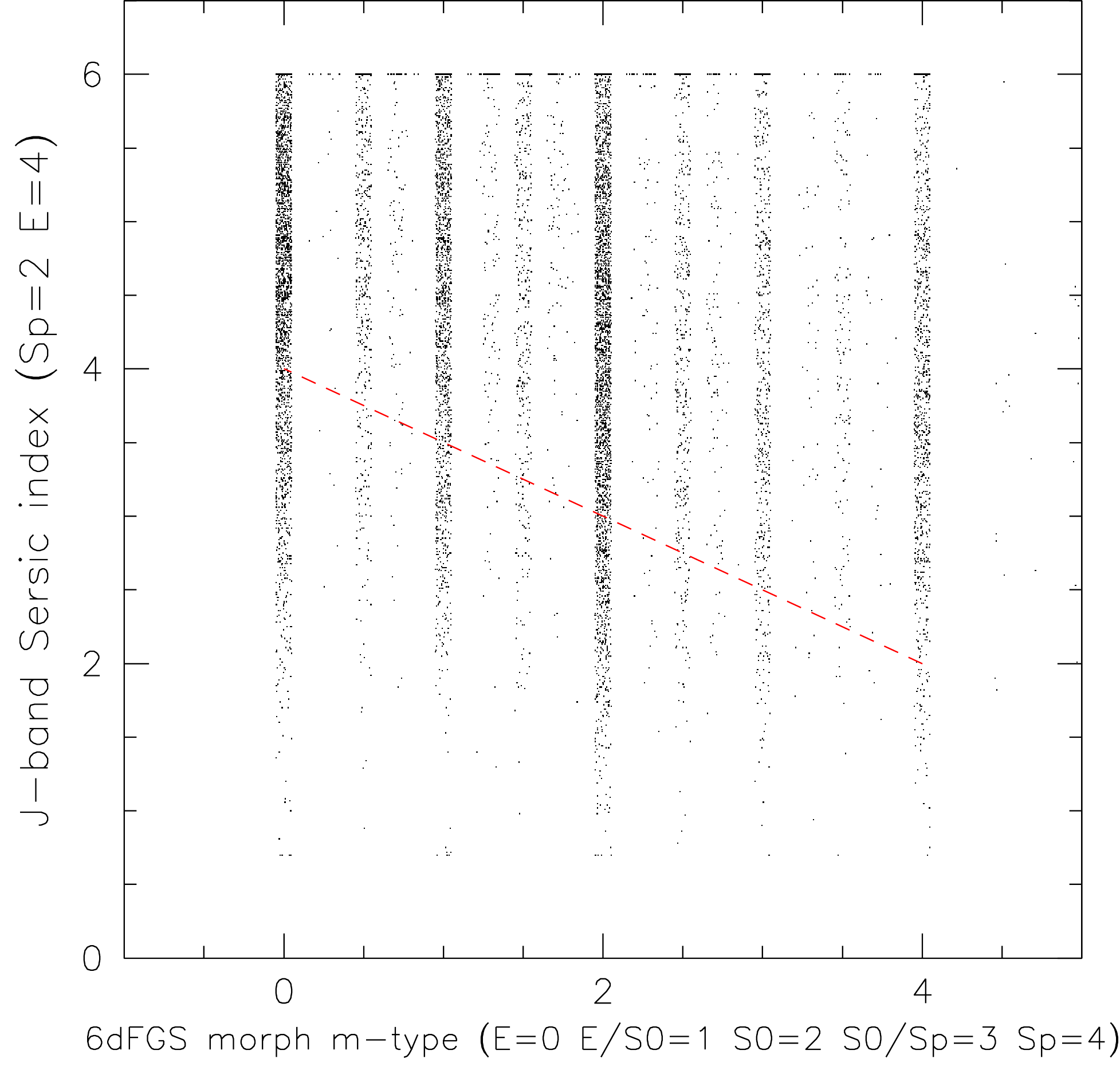} 
\caption{Comparison of visual morphological 
classification and $J$-band S\'{e}rsic. The trend
of S\'{e}rsic index of 4 for ellipticals to 2 for
spirals (bulges) is shown as the dashed line.
}
\label{fig:sersic_type}
\end{figure}

As an alternative indicator of morphological type, we considered using
our GALFIT-derived $J$-band S\'{e}rsic 
indices (see Table~\ref{tab:photo_data}). While there is a
trend of decreasing S\'{e}rsic index as the morphogical type
changes from elliptical to spiral, the scatter is large 
(see Figure~\ref{fig:sersic_type}).
The low signal-to-noise of the 2MASS image data
and the bulge-dominated nature of the NIR light makes these
2MASS-based S\'{e}rsic indices a poor discrimator of morphogical type 
and therefore the indices were not used for the selection of our FP sample.

All galaxies were also assessed according to three quality controls and
a code was assigned in each case (Table \ref{tab:other}). The three
aspects were as follows.
\begin{enumerate}
\item {
\it Fibre coverage of the galaxy:} whether the fibre aperture was
  dominated by light from the bulge alone, a disk (with little or no
bulge), or a combination of the two.
By {\it dominated} we mean that the estimated bulge contribution
to the observed spectrum was $>$80\%.
By {\it little or no} we mean that the bulge contribution 
was estimated at $<$20\%. 

  Figure~\ref{figmorph0} shows an
  example of a nearby barred-spiral galaxy where the bulge fills the
  fibre aperture, giving an early-type spectrum for a late-type galaxy.
\item {\it Source confusion:} flagging cases of galaxy/galaxy or
  galaxy/star overlap (within or adjacent to the fibre aperture), or
  other special cases of note (e.g.\ Galactic sources, high redshift
  QSOs).
\item {\it Spectral nature:} broadly characterising whether the spectrum
  was dominated by absorption or emission-line features, or both. In
  addition, at this stage redshifts were checked for gross errors (none
  were found).
\end{enumerate}

\begin{figure}
\includegraphics[clip,width=\plotwidth]{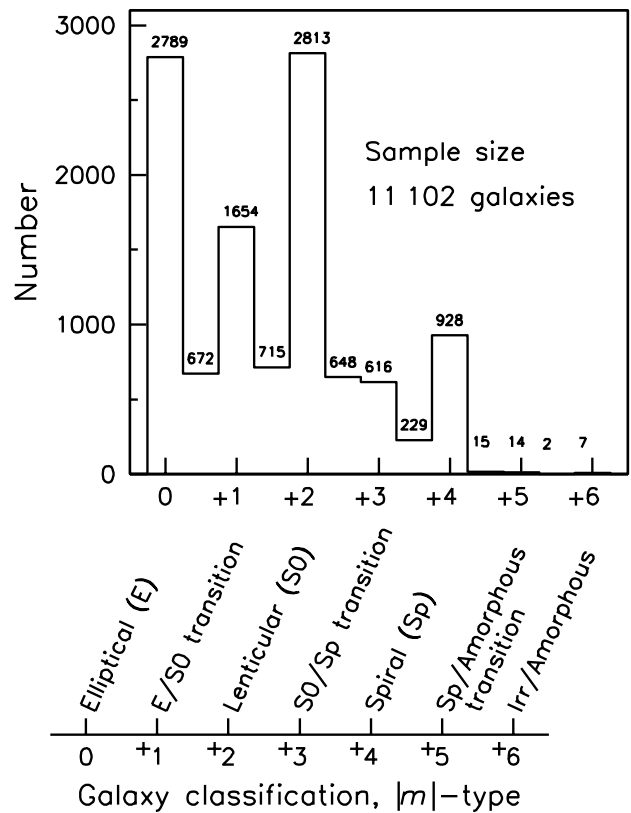} 
\caption{Distribution of galaxy types assigned to the 6dFGSv sample,
  binned by $|m|$-type (i.e.\ the absolute value of $m$-type). The
  lower axis gives the relationship between $|m|$-type and morphology. 
  The counts at the half-classification step arise due to the effect of 
  averaging integer classifications over multiple observations.
}
\label{figmorph2}
\end{figure}

\begin{figure}
\includegraphics[clip,width=\plotwidth]{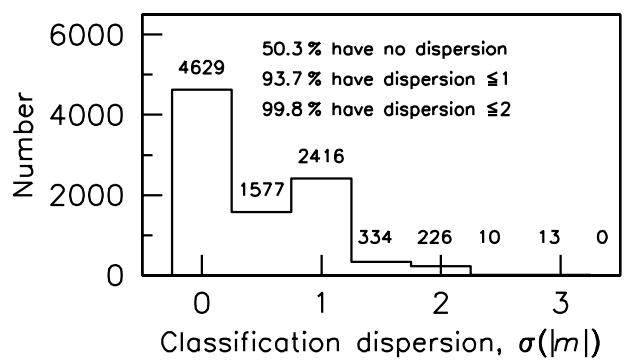} 
\caption{Distribution of the dispersion in the $|m|$-type values
  assigned independently. The subsample shown excludes the 1897 galaxies
  with only a single classification, which have $\sigma = 0$ by
  default.}
\label{figmorph3}
\end{figure}

\subsection{Results}

Figure~\ref{figmorph2} summarises the results of our morphological
classification for the full 6dFGSv sample. The final $m$-type for a
given galaxy was the average of the individual values assigned by the
classifiers for that galaxy. The number of galaxies for each type are
shown, divided into bins of $|m|$-type, which has the aforementioned
property of merging the edge-on cases with all other disk galaxies.
There were 1712 galaxies (15\% of the sample) classified as edge-on by
at least half of the people viewing each one.

The sample was dominated by elliptical and S0 types (9\,291 galaxies
with $|m|$$<$$+3$, or 84\% of the full sample). This was not unexpected,
since the input sample was selected by spectroscopic matching to
early-type galaxy template spectra. Importantly, Figure~\ref{figmorph2}
also shows a small but non-negligible number of spirals (928 galaxies or
8\%). This was also not surprising, since these are cases where the
bulge of the spiral fully fills the fibre aperture. As noted above,
fibre coverage was one of the quality control parameters noted for each
galaxy independently of morphology (Table~\ref{tab:other}).
Figure~\ref{figmorph2} also shows that transitional cases are relatively
few. We do not believe this reflects the true occurrence of these
systems, as the effect of increased distance blurs the distinction
between different galaxy types.

The simplest internal consistency check we used was to examine the
dispersion, $\sigma(|m|)$, in $|m|$-types assigned by all classifiers
for a given galaxy. Cases of full agreement between classifiers should
result in zero dispersion. Figure~\ref{figmorph3} shows the number of
galaxies in bins of $\sigma(|m|)$, excluding the 1\,897 with only one
classification. Over 93\% of the sample has a classification dispersion
of less than a single class ($\sigma(|m|) \leq 1$), including 4\,629
galaxies with zero dispersion. A dispersion of less than two classes
($\sigma(|m|) \leq 2$) encompasses 99.8\% of the sample. Given that the
major types (elliptical/S0/spiral) are all separated by $|m|$-type
intervals of $+2$, Figure~\ref{figmorph3} demonstrates our
self-consistency in distinguishing the major types from one another for
the vast majority of the 6dFGSv sample.

The effects of increasing distance complicated the task of galaxy
classification in two ways. We divide these into imaging and
spectroscopic classification biases.
 
{\it Imaging classification bias} relates to the increasing difficulty
of recognising a galaxy type at the sensitivity and spatial resolution
limits of the system. Faint and detailed structures such as spiral arms
and dust lanes demand the most stringent imaging requirements. As a 
consequence, galaxy classification schemes based purely on imaging tend to
increasingly misclassify late-type galaxies as early-types at higher
redshifts. This effect is evident in large catalogues such as Galaxy Zoo
\citep[e.g. ][]{bamford09}, where the number of unequivocal spirals (in
the sense that more than 80\% of the classifiers agreed) seen at
redshift $z \sim 0.08$ was half the number seen at $z \sim 0.01$, while
over the same interval the number of ellipticals doubles. Obviously this
is a major concern for surveys considering morphology in the absence of
other (e.g.\ spectral) information; fortunately, this was not the case
for 6dFGSv.
 
{\it Spectroscopic classification bias} arises when galaxy types are
assigned on the basis of their spectra through a fixed aperture. Such
apertures are almost always circular and positioned over the galaxy
centre, meaning that light from the outer regions is only sampled in
cases where the galaxy is sufficiently small and/or distant. This means
that a sample ostensibly of early-types can in fact contain late-type
galaxies whose proximity ensures the bulge fills the fibre aperture,
producing an early-type spectrum. Clearly this effect is
distance-dependent.

Both forms of bias act to underestimate the numbers of spirals in
different redshift regimes. Spectroscopic classification bias operates
at lower redshifts while imaging classification bias tends to be an
issue at higher redshifts. However, since we were concerned with
constructing a FP sample of early types, biases that act to miss late
types were not important. The only effect we concerned ourselves with
were late types being mistaken for early types as a consequence of
spectroscopic classification bias. This bias was present in 6dFGSv but
was also readily identifiable: our visual screening of the galaxy images
ensured that spiral bulges were identified correctly. As the 6dFGSv
sample was limited to redshifts $z < 0.055$, image classification bias
was limited, as experience showed that reliable classification is
possible to an apparent magnitude-limit of $K \sim 13$ (or equivalently,
$b_{\rm J} \sim 17$). A more stringent test was to examine the
composition of our sample with increasing distance.

\begin{figure}
\includegraphics[clip,width=\plotwidth]{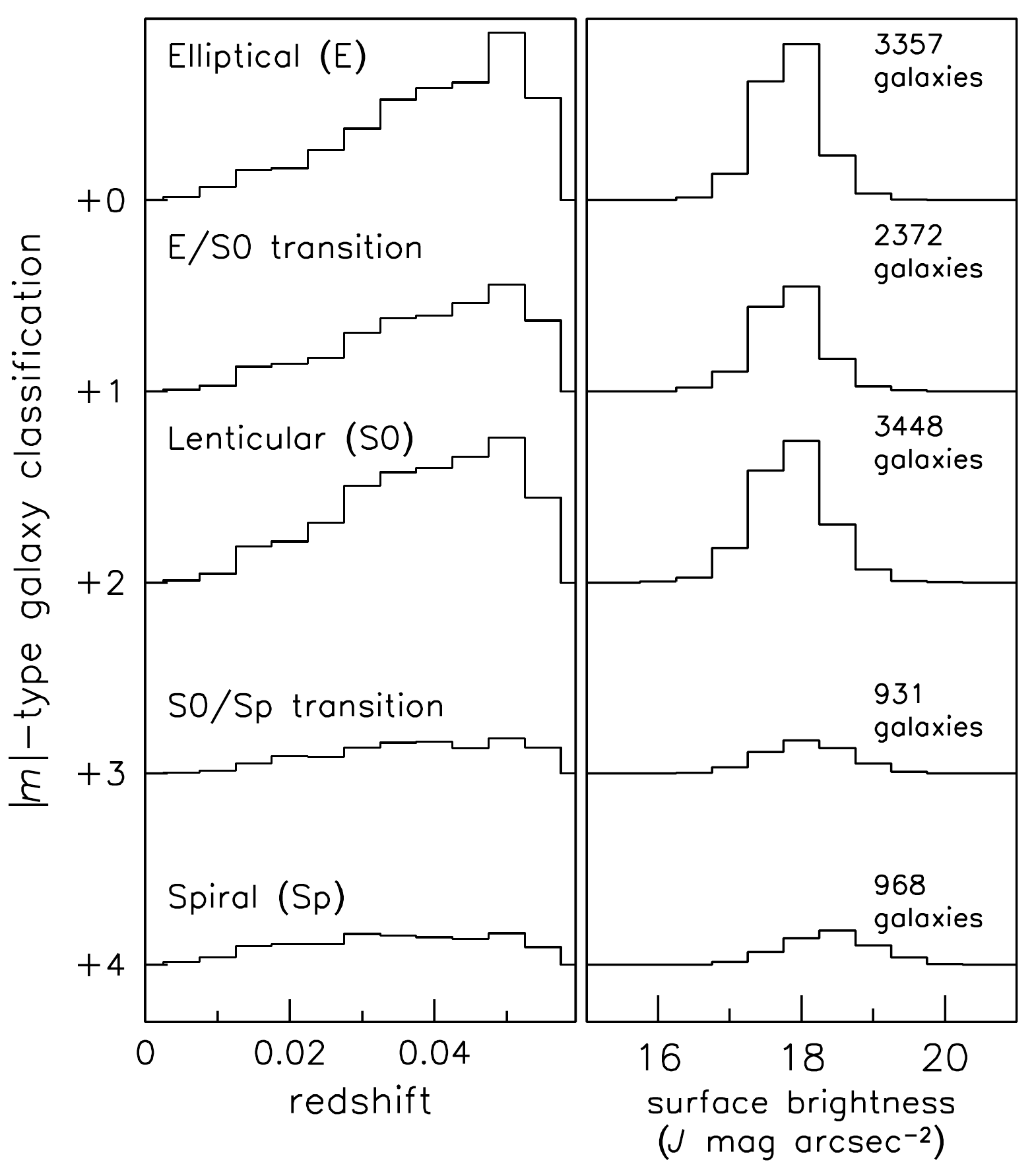} 
\caption{Galaxy distribution in each $|m|$-type classification bin as a
  function of redshift ({\it left}) and surface brightness ({\it right}).}
\label{figmorph4}
\end{figure}

These effects were explored by measuring the distribution of types in
terms of surface brightness and redshift. The surface brightness
distributions in Figure~\ref{figmorph4} tended toward fainter values for
spirals and spiral transition cases. This was not unexpected given the
initial preselection of our sample based on bulge spectral properties.
At a fixed luminosity, a spiral galaxy will, due to its disk, have a
lower overall surface brightness than an elliptical. The fact that the
range of surface brightness seen in Figure~\ref{figmorph4} does not
change more was likely due to our preference for early-types where total
galaxy luminosity was dominated by the contribution from the bulge over
the light from any disk.

\begin{figure}
\includegraphics[clip,width=\plotwidth]{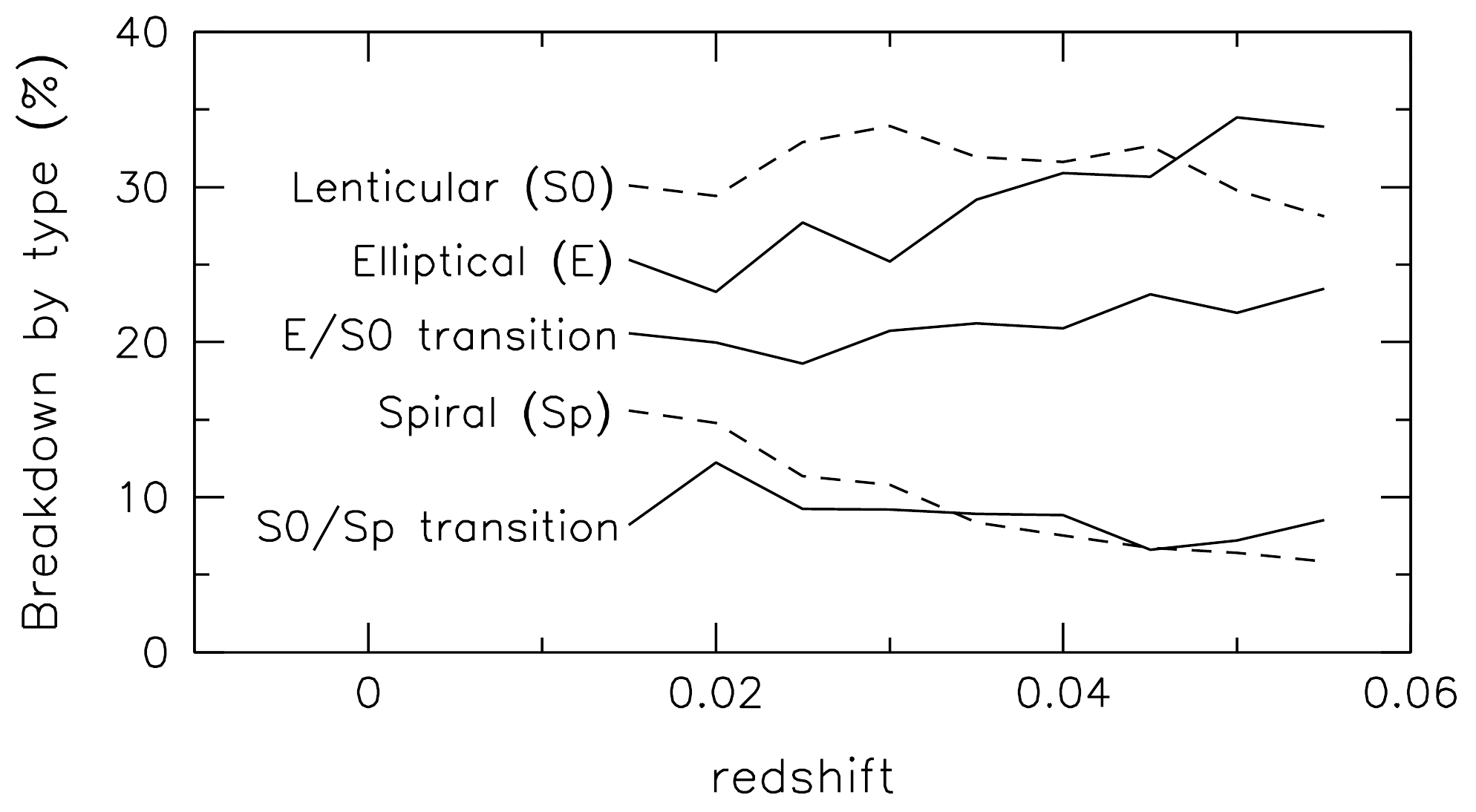} 
\caption{Relative galaxy number as a function of redshift.}
\label{figmorph5}
\end{figure}

Figure~\ref{figmorph5} shows that the mix of types found by our
classification scheme did not change appreciably over the redshift range
considered. We recorded a small rise in the fraction of ellipticals
(from 25\% to 35\%) while the numbers of S0s and transition cases
remained flat. The corresponding decline in the number of spirals (from
15\% to 6\%) we attribute to two effects. First, the fact that the
fraction of early types increases with luminosity means that they will
also increase with redshift in a magnitude-limited sample. Secondly, the
spectroscopic preselection (and the fixed fibre aperture employed) will
include the very lowest redshift spirals because of the early-type
appearance of their bulge-dominated spectra. At higher redshifts, the
spirals are spectroscopically rejected as more of the fibre aperture
flux is due to the light of the star-forming disks, including emission
lines from the H{\sc II} regions. This emphasises the importance of our
visual screening to counter the spectroscopic classification bias
inherent in the input sample.

The other check we used was to benchmark our classifications against
those of another group. We made a comparison between Galaxy Zoo
morphological types and our assignments for the sample of 281 galaxies
common to both. Galaxy Zoo \citep{lintott08} is the compilation of basic
morphological typing obtained by crowd-sourcing the visual
classification of 893\,212 galaxies from the Sixth Data Release of the
Sloan Digital Sky Survey \cite[DR6;][]{adelmanmccarthy08}. For the
Galaxy Zoo sample there was a median of 34 independent views per galaxy
using elliptical and spiral as the two main classification divisions.

\begin{figure}
\includegraphics[clip,width=\plotwidth]{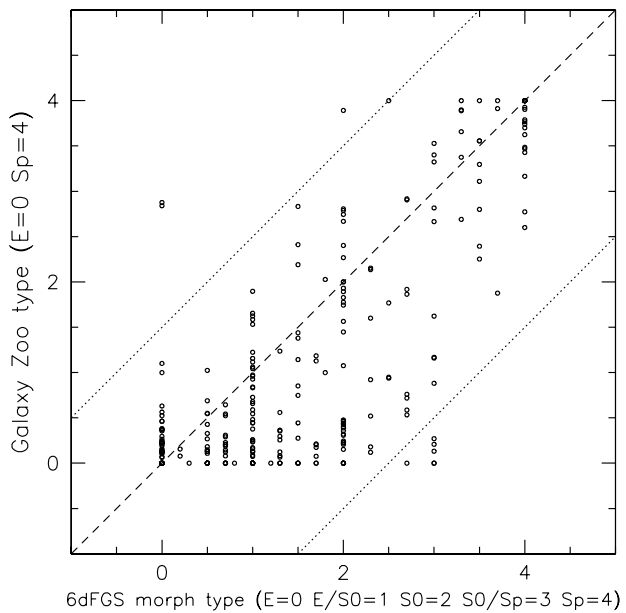} 
\caption{Comparison of Galaxy Zoo \citep{lintott08} and 6dFGS
  morphological types for a sample of 281 galaxies in common. The Galaxy
  Zoo classification has been transformed to the 6dFGS scheme indicated
  on the horizontal axis. The dashed line shows the 1:1 relation and the
  dotted lines show the $(+1.5,-2.5)$ boundaries that encompass most of
  the scatter. Galaxy Zoo classifications have a tendency towards
  earlier types.}
\label{figmorph6}
\end{figure}

Figure~\ref{figmorph6} shows the comparison between the 6dFGSv and
Galaxy Zoo. To facilitate comparison we rescaled the combined Galaxy Zoo
type on an equivalent scale of 0 to 4 (elliptical to spiral). Edge-on
and problem cases have been excluded from the sample, as have galaxies
with only one independent classification. The agreement between 6dFGSv
and Galaxy Zoo is good with minimal scatter ($\sigma(|m|) \sim 1$). The
amount of scatter about the 1:1 relation is less than the size of the
step ($\Delta|m| = 2$) separating the major galaxy classes (E:S0:Sp).

However, Figure~\ref{figmorph6} also shows a tendency towards
early-typing by Galaxy Zoo relative to the 6dFGSv classes, in that a
greater number of galaxies lie below the 1:1 line. We contend that this
was due to the redder optical passbands used by Galaxy Zoo, and their
reliance on composite images rather than examination of separate images
in each passband. Morphological typing of 6dFGSv galaxies was done using
multiple independent optical images ($b_{\rm J}r_{\rm F}$), with
supporting near-infrared $JHK$ frames, and a colour near-infrared
composite. In contrast, Galaxy Zoo classifications used only a
single-colour composite image made from optical $gri$ passbands. Since
none of these sample the galaxy light shortward of $\lambda \lesssim
4000$\,\AA, the blue light of the OB stars that dominate the arms of
spiral galaxies is under-represented in the composite frames. Therefore,
galaxies with fainter spiral features tend to be classified as S0s
(predominantly on the basis of their bulge).

The morphological classifications of 6dFGSv were used for both trimming
problem galaxies as well as examining FP trends with morphology
\citep{magoulas12}. The characteristics of this morphologically clean
sample are discussed below.

\begin{table}
\begin{center}
\caption{6dFGS quality control classification codes for (1)~fibre
  coverage, (2)~source confusion, and (3)~spectral features.
  \label{tab:other}}
\begin{tabular}{lcl}
\hline
Fibre coverage & Code & Description \\
\hline
& & \\
bulge & 1 
& Bulge is present (S0-Sp) and it \\
& & fills the fibre aperture, or the\\
& & galaxy has no disk (E).\\
& & \\
disk & 2 
& Galaxy shows no bulge (Irr), or \\
& & fibre misses bulge completely.\\
& & \\
bulge$+$disk & 3 
& Fibre captures light from both \\
& & bulge and disk, or the galaxy is \\
& & an edge-on disk.\\
 & & \\
\hline
Source confusion & Code & Description \\
\hline
& & \\
none & 1 & --- \\
& & \\
galaxy$+$star & 2 
& The light in the fibre has contri- \\
& & butions from a nearby star.\\
& & \\
galaxy$+$galaxy & 3 
& The light in the fibre has contri- \\
& & butions from two or more galaxies.\\
& & \\
not a galaxy & 4 
& The fibre light is dominated by a \\ 
& & star/PN/HII region/QSO.\\
& & \\
other & 5 
& Any problem not fitting the above.\\
 & & \\
\hline
Spectral features & Code & Description \\
\hline
& & \\
absorption only & 1 
& Galaxy continuum with only clearly \\
& & discernible absorption features.\\
& & \\
absorption$+$emission & 2
& Both absorption and emission features \\
& & clearly discernible above the noise.\\
& & \\
emission only & 3 
& Little or no detected continuum and \\
& & no discernible absorption features, \\
& & or absorption features of much lower \\
& & prominence than emission lines. \\ 
& & \\
\hline
\end{tabular}\\
\end{center}
\end{table}

\section{6\lowercase{d}FGS\lowercase{v} Fundamental Plane Sample}
In this section we describe how the 6dFGSv Fundamental Plane sample and
catalogue is constructed from the measurements described in previous
sections. 

\subsection{Velocity dispersion aperture correction}

Our measured velocity dispersions ($\sigma$) are corrected to a standard
physical aperture size of ${R_{e}}/{8}$, using the empirically-derived
formula given by \citet{jorgensen95b}:
\begin{equation}
\frac{\sigma_{0}}{\sigma} =
\left(\frac{R_{e}/8}{R_{ap}}\right)^{-0.04}
\end{equation}
where $\sigma_{0}$ is the `central' velocity dispersion within a
circular aperture of radius ${R_{e}}/{8}$. 
The 6dF fibre diameter is 6.7\,arcsec and so $R_{ap} = 3.35$\,arcsec. 
Rather than use the above equation directly 
with our NIR-based $R_{e}$ measurements, 
we first use an empirical 
relation that maps  $R_{e}(NIR)$ to $R_{e}(optical)$
in order to provide a 
closer match to previous optical-based studies.

Using a sample of galaxies in common with the
ENEAR survey \citep{alonso03}, we derived the following relation between
the $R$- and $J$-band effective radii
\begin{equation}
\log R_e^R = 1.029 \log R_e^J + 0.140 ~.
\end{equation}
The range of $R_e^J$ for our sample is 1.6 to 16\,arcsec, corresponding
to an $R_e^R$ range of 2.2 to 24\,arcsec. The correction factor to the
measured velocity dispersion ranges from 1.107 to 1.010.

\subsection{\boldmath$R_e$ and \boldmath$\langle I_e \rangle$}

We convert the effective radii $R_e$ in arcseconds, as given in
Table~\ref{tab:photo_data}, to effective radii $R_{e}^{kpc}$ in physical
units of $h^{-1}$\,kpc using
\begin{equation}
\log R_{e}^{kpc} = \log R_{e} + \log D_{A} + \log{(1000/206265)}
\label{eq:physrad}
\end{equation}
where $D_{A}$ is the galaxy's angular diameter distance in units of
$h^{-1}$\,Mpc derived from the measured galaxy redshift (in the local
CMB rest frame) using a flat cosmology with $\Omega_{m}=0.3$,
$\Omega_{\Lambda}=0.7$ and $H_0=100\,h$\,km\,s$^{-1}$\,Mpc$^{-1}$.

We find that 3186 of the galaxies in the sample are members of groups,
as identified by the friends-of-friends grouping algorithm outlined by
\citet{magoulas12}. For these galaxies, we use the redshift of the
group, rather than the individual galaxy redshift, to compute the
redshift distance and the corresponding physical radius. The group
redshift is defined as the median redshift of all galaxies in the group.

Corrections were made to the values of $\langle \mu_{e} \rangle'$ for
(i)~the cosmological surface brightness dimming, (ii)~the k-correction,
using the approximations of +0.7z, +0.2z and $-$3.3z mag for the $J$-,
$H$- and $K$-bands respectively \citep{persson79}, and (iii)~galactic
extinction using the values given by \citet{schlafly11} as report by 
NASA NED. The fully
corrected effective surface brightness is calculated as
\begin{equation}
\langle \mu_{e} \rangle = \langle \mu_{e} \rangle' -
2.5 \log{(1+z)^4} - k_\lambda - A_\lambda
\end{equation}
where we use the redshift in the local CMB rest frame.

It is most natural to have all FP parameters in logarithmic units, so
surface brightness values were converted from magnitude units (i.e.\
\sbmag\ in mag\,arcsec$^{-2}$) to log-luminosity units (i.e.\ \sblum\ in
$\lsun$\,pc$^{-2}$) using
\begin{equation}
\log \langle I_{e} \rangle = 0.4 {\rm M}_\odot^\lambda 
- 0.4 \langle \mu_{e} \rangle + 2 \log{(206265/10)}
\end{equation}
where the absolute magnitude of the Sun, ${\rm M}_\odot^\lambda$,
depends on the passband. For the $J$-band, ${\rm M}_\odot^J$=3.67; for
the $H$-band, ${\rm M}_\odot^H$=3.33; and for the $K$-band, 
${\rm M}_\odot^K$=3.29; these values for the absolute magnitude of the
Sun are from http://mips.as.arizona.edu/$\sim$cnaw/sun.html.

\subsection{Construction of the 6dF Fundamental Plane Sample}

By combining both the velocity dispersion measurements from Table
\ref{tab:sigma_data} (with the above aperture correction) and 2MASS
photometric measurements from Table \ref{tab:photo_data} (with the
corrections and conversion given above) we construct the 6dF FP
sample.

It is important to distinguish between four overlapping samples used in
our analysis. There are the three samples used to fit the FP, i.e.\ the
$J$-, $H$- and $K$-band FP samples, while the fourth is the $J$-band
sample used for the peculiar velocity analysis. Details of the galaxy
selection for the latter are described in detail in \citet{magoulas12},
and only a brief summary is given here.

In Section~2.4 we described an initial set of selection criteria on the
velocity dispersion sample ($S/N > 5$, $R > 8$, $\sigma > 100$\kms
~that resulted in a sample of 11\,315 galaxies. In Section~3, we reported
photometric parameters for 11\,102 of these galaxies. This is the parent
sample from which additional selection cuts were made to produce a
sample suitable for the FP analysis. These additional cuts are listed in
\citet{magoulas12} Table~1. We first limited the sample to galaxies with
$3\,000 \le cz_{CMB} \le 16\,120$\kms\ and $\log{\sigma_0} \ge 2.05$
(this latter cut is imposed rather than the original $\log{\sigma_0} = 2.00$
to ensure a uniform limit after the aperture correction was applied). We
also eliminated galaxies on the basis of the visual classification
of the morphology described in Section~4; galaxies were excluded if at
least one of four conditions was met: 
(i)~galaxy morphology classified as irregular or amorphous;
(ii)~galaxy identified as edge-on with a full dust lane;
(iii)~significant fraction ($>$80\%) of the light in fibre is from the
disk; or (iv)~light in fibre is contaminated by nearby star, galaxy or
defect.

These selection cuts resulted in a sample of 9\,656 galaxies. From that
sample, we constructed individual $J$-, $H$-, and $K$- band samples by
imposing additional apparent magnitude cuts in each of the three bands.
Our sample had slightly brighter flux limits than the original 6dFGS
magnitude limits \citep{jones09}, reflecting the changes in the 2MASS
(and, consequently, 6dFGS) magnitude limits that occurred after the
6dFGS sample was selected. To maintain high completeness in each
passband over the whole sample area, we imposed magnitude limits of 
$J \leq 13.65$, $H \leq 12.85$ and $K \leq12.55$.

In our FP fitting procedure, each galaxy was weighted by the inverse of
its selection probability. This means galaxies are weighted by $1/S$,
where $S$ is the fraction of the survey volume in which a galaxy of that
absolute magnitude could have been included in the sample. For each of
the three photometric bands, we excluded galaxies for which $S < 0.05$
in order to prevent a small number of heavily weighted galaxies from
strongly biasing the FP fit. \citet{magoulas12} also defined a $\chi^2$
statistic, computed for each galaxy, which measured the galaxy's
deviation from the model FP relation. Galaxies with $\chi^2$ values
greater than 12, representing a deviation from the model of more than
about $3.5\sigma$, were also eliminated from the sample.

After applying these selection criteria, we obtained, $J$-, $H$-, and
$K$-band samples of 8\,803, 8\,472, and 8\,461 galaxies respectively.
However, once the FP has been fit, there are various
FP applications for which it is appropriate to include some of the
discarded galaxies, e.g.\ galaxies excluded by the lower redshift limit
of $cz_{CMB}\,=\,3000$ km\,s$^{-1}$ and the selection probability limit
of $S=0.05$. In \cite{springob14}, for example, we present the
derivation of galaxy distances and peculiar velocities for the 6dFGSv
sample. This is the sample of galaxies for which we have derived
distances and peculiar velocities from the $J$-band data, and it
includes these low-redshift and low-selection-probability galaxies,
resulting in a total of 8\,885 objects in the peculiar velocity sample.
The parameters derived for each galaxy in these overlapping samples are
summarised in Table~\ref{tab:6dfcat}; the data table is given as
Table~\ref{tab:fpdata}.

\begin{table*}
\begin{center}
\caption[6dFGS Fundamental Plane sample parameters]{Summary of 6dFGS
  Fundamental Plane sample parameters. For each parameter, the table
  below lists its name, units, and a short description. Units enclosed
  by brackets are logarithmic quantities. 
} 
{\footnotesize
\begin{tabular}{@{}cllll}
\hline
Table Column & Parameter & Unit & Description\\
\hline
(1)  & 6dFGSid   & ---    & Source name in 6dFGS catalogue \\
(2)  & 2MASSid   & ---    & Source name in 2MASS XSC catalogue \\
(3)  & R.A.      & degree & Right ascension (J2000) \\
(4)  & Dec.      & degree & Declination (J2000) \\
(5)  & $cz_{CMB}$ & \kms   & CMB frame galaxy redshift \\
(6)  & $\log{R_e^{kpc}}(J)$          & [kpc\,$h^{-1}$] & Logarithm of effective radius in the $J$-band \\
(7)  & $\epsilon_{\log{R_e^{kpc}}(J)}$ & [kpc\,$h^{-1}$] & Error in $\log{R_e^{kpc}}(J)$ \\
(8)  & $\log{R_e^{kpc}}(H)$          & [kpc\,$h^{-1}$] & Logarithm of effective radius in the $H$-band \\
(9)  & $\epsilon_{\log{R_e^{kpc}}(H)}$ & [kpc\,$h^{-1}$] & Error in $\log{R_e^{kpc}}(H)$ \\
(10) & $\log{R_e^{kpc}}(K)$          & [kpc\,$h^{-1}$] & Logarithm of effective radius in the $K$-band \\
(11) & $\epsilon_{\log{R_e^{kpc}}(K)}$ & [kpc\,$h^{-1}$] & Error in $\log{R_e^{kpc}}(K)$ \\
(12) & \sblum\,$(J)$                          & [$\lsun$\,pc$^{-2}$ ] & Logarithm of mean $J$ band surface brightness within $R_e(J)$ \\
(13) & $\epsilon_{\log \langle I_{e} \rangle (J)}$ & [$\lsun$\,pc$^{-2}$ ] & Error in  \sblum\,$(J)$ \\
(14) & \sblum\,$(H)$                          & [$\lsun$\,pc$^{-2}$ ] & Logarithm of mean $H$ band surface brightness within $R_e(H)$ \\
(15) & $\epsilon_{\log \langle I_{e} \rangle (H)}$ & [$\lsun$\,pc$^{-2}$ ] & Error in  \sblum\,$(H)$ \\
(16) & \sblum\,$(K)$                          & [$\lsun$\,pc$^{-2}$ ] & Logarithm of mean $K$ band surface brightness within $R_e(K)$ \\
(17) & $\epsilon_{\log \langle I_{e} \rangle (K)}$ & [$\lsun$\,pc$^{-2}$ ] & Error in  \sblum\,$(K)$ \\
(18) & \vd                     & [\kms ] & Logarithm of aperture-corrected central velocity dispersion \\
(19) & $\epsilon_{\log{\sigma_0}}$ & [\kms ] &  Error in $\log\sigma_0$ \\
(20) & $m$-type   & --- & Average morphological type classification, see Table~\ref{tab:morph} \\
(21) & GroupID    & --- & Group or cluster identification number \\
(22) & $N_r$      & --- & Richness of galaxy group or cluster \\
(23) & $cz_{group}$ & \kms  & Group or cluster median CMB redshift \\
(24) & $d_5$      & Mpc $h^{-1}$            & Projected comoving distance to the $5^{\rm th}$ nearest neighbour (not available for 1165 galaxies) \\
(25) & $\Sigma_5$ & gals\,Mpc$^{-2}$ $h^{2}$ & Surface density measured to the $5^{\rm th}$ nearest neighbour (not available for 1165 galaxies) \\ 
(26), (28), (30) & Sample Code & - & The selection criteria that a galaxy satisfies are encoded in this sample code ($JHK$). These codes are \\
     &           &             & six-digit binary strings, where digit\,=\,0 or~1 indicates that the galaxy fails or satisfies the selection criterion.\\
     &           &             & The six selection criteria (in left to right string order) are: (1)~\vd\ $> 2.05$; (2)~$cz > 3000$ and \\
     &           &             & $cz < 16120$\,km\,s$^{-1}$; (3)~morphological classification acceptable; (4)~apparent magnitude $< 13.65$ ($J$), \\
     &           &             & 12.85 ($H$) or 12.55 ($K$); (5) selection probability $> 0.05$ ($JHK$); (6) $\chi^2 < 12$ ($JHK$). \\
(27), (29), (31) & PV Code & - & Code to indicate whether galaxy excluded from (0) or included in (1) the $JHK$ 6dFGSv sample. \\
\hline
\label{tab:6dfcat}
\end{tabular}
}
\end{center}
\end{table*}

\begin{table*}
\caption{The 6dFGS Fundamental Plane Catalogue. Table 7 provides a
  description of each column. For ease of presentation here, the
  columns here are wrapped over three lines. 
  Missing or not applicable values are set to --1, e.g. in column (21) for 
  galaxies not assigned to a group.
  The full version of this table, with one galaxy per row, 
  is provided in the online Supporting Information.
}
  
\label{tab:fpdata}
\begin{tabular}{lrrrrrrrrrrrr}
\hline \hline
\multicolumn{1}{c}{(1)}  &  
\multicolumn{1}{c}{(3)}  &
\multicolumn{1}{c}{(4)}  &
\multicolumn{1}{c}{(5)}  & 
\multicolumn{1}{c}{(6)}  &
\multicolumn{1}{c}{(7)}  & 
\multicolumn{1}{c}{(8)}  &  
\multicolumn{1}{c}{(9)}  &
\multicolumn{1}{c}{(10)} &
\multicolumn{1}{c}{(11)} \\
\multicolumn{1}{c}{(2)}  & 
\multicolumn{1}{c}{(12)} &
\multicolumn{1}{c}{(13)} & 
\multicolumn{1}{c}{(14)} &
\multicolumn{1}{c}{(15)} & 
\multicolumn{1}{c}{(16)} &  
\multicolumn{1}{c}{(17)} &
\multicolumn{1}{c}{(18)} &
\multicolumn{1}{c}{(19)} &  
\multicolumn{1}{c}{(20)} \\ &
\multicolumn{1}{c}{(21)} &  
\multicolumn{1}{c}{(22)} & 
\multicolumn{1}{c}{(23)} &
\multicolumn{1}{c}{(24)} &
\multicolumn{1}{c}{(25)} & 
\multicolumn{1}{c}{(26)} &
\multicolumn{1}{c}{(27)} & 
\multicolumn{1}{c}{(28)} &  
\multicolumn{1}{c}{(29)} &
\multicolumn{1}{c}{(30)} &
\multicolumn{1}{c}{(31)} \\
\hline
g0000144-765225          & 0.00399 & --76.87364 & 15941  & 0.465 & 0.057 & 0.363 & 0.069  & 0.261 & 0.092 \\
2MASXJ00001440-7652248   & 2.940 & 0.084 & 3.266   & 0.102 & 3.489 & 0.136 & 2.137 & 0.073 & 2.2\\
                         & --1 & --1 & --1 & 4.2758 &  0.1176 & 111111 & 1 & 111111 & 1 & 111111 & 1 
\vspace{0.5mm}\\
g0000222-013746          & 0.00615 & --1.62947  & 11123  & 0.024 & 0.048 & --0.010 & 0.057 & --0.087 & 0.079 \\
2MASXJ00002213-0137463   & 3.708 & 0.071 & 3.921 & 0.085 & 4.067 & 0.118 & 2.338 & 0.023 & 2.5 \\
                         & --1 & --1 & --1 & 6.4318 & 0.0385 & 111111 & 1 & 111111 & 1 & 111111 & 1
\vspace{0.5mm}\\
g0000235-065610          & 0.00652  & --6.93619 & 10920  & 0.343 & 0.047  & 0.370 & 0.057  & 0.133 & 0.083 \\
2MASXJ00002348-0656103   & 3.067 & 0.070 & 3.161 & 0.084 & 3.562 & 0.123 & 2.219 & 0.045 & 3.7 \\
                         & --1  & --1  &   --1 &  6.2142 & 0.0448 & 111111 & 1 & 111111 & 1 & 111111 & 1
\vspace{0.5mm}\\
g0000251-260240          & 0.00697 & --26.04450 & 14926  & 0.371 & 0.046  & 0.338 & 0.055  & 0.310 & 0.074 \\
2MASXJ00002509-2602401   & 3.328   & 0.068 & 3.531 & 0.082 & 3.592 & 0.110 & 2.369 & 0.040 & 2.3 \\
                         & --1  & --1 &  --1  & 3.2827 & 0.1434 & 111111 & 1 & 111111 & 1 &  111111 &  1
\vspace{0.5mm}\\
g0000356-014547          &  0.00990 & --1.76317 & 6956  & 0.316 & 0.042 & 0.286 & 0.052  & 0.320 & 0.067 \\
2MASXJ00003564-0145472   &  2.859 & 0.062 & 3.029 & 0.077 & 3.040 & 0.099 & 2.144 & 0.066 & 3.0 \\
                         & --1  & --1 & --1   &  3.8994 & 0.1036 & 111111 & 1 &  111111 & 1 &  111111 & 1\\

\hline
\end{tabular}
\end{table*}

Most of the parameters in this table have already been defined in this
paper, however we elaborate here on Columns~(21)--(31). As explained in
Section~5.2, we make use of a group catalogue explained in
\citet{magoulas12}, which is drawn from the much larger set of all
galaxies with redshifts in 6dFGS, not just those in the FP sample.
Column~(21) is the group ID number for the galaxy in question, in those
cases for which the galaxy is in a group. We also define $N_r$ in Column
(22) as the group richness. Because the faintest members of a group
might not have been observed, the richness of a group is defined as the
number of observed galaxies in the group brighter than a specified
absolute magnitude, chosen so that galaxies brighter than this would be
visible throughout the sample volume. Column~(23) is the redshift of the
group, taken as the median redshift of all galaxies in the group
(including those within the 6dFGS redshift sample that are not in the FP
subsample).

We also include two additional measures of local environment.
Column~(24) is the projected comoving distance to the 5th-nearest
neighbor ($d_5$) in $h^{-1}$\,Mpc. Column~(25), the surface density
($\Sigma_5$), is defined using the projected co-moving distance to the
5th nearest neighbour within $\pm$1000\,km\,s$^{-1}$ within a volume
limited density-defining population with $M_{K}\,<$ --22 mag. Further
details of $\Sigma_5$ can be found in \citet{brough13}.

Columns~(26) to (31) are binary codes to denote whether or not a galaxy
is included in the various subsamples previously described. For
inclusion in a sample six selection criteria are assessed. These are
described in Table~\ref{tab:6dfcat}. 
Values in column~(26) refer to the $J$-band data
and are a six-character sequence which indicates which of the six
selection criteria are met; 1 denotes inclusion and 0 denotes exclusion.
Only galaxies with a value of 111111 in column~(26) are included in the
final $J$-band sample. Similarly, columns~(28) and (30) give the six-character
codes for the $H$-band and $K$-band data respectively. Columns~(27),
(29) and~(31) denote whether or not a galaxy is included in the $J$-,
$H$-, $K$-band 6dFGSv peculiar velocity samples respectively. As
previously noted, we primarily used the $J$-band 6dFGSv sample for our
peculiar velocity studies.

\begin{figure}
\centering
\includegraphics[width=0.45\textwidth]{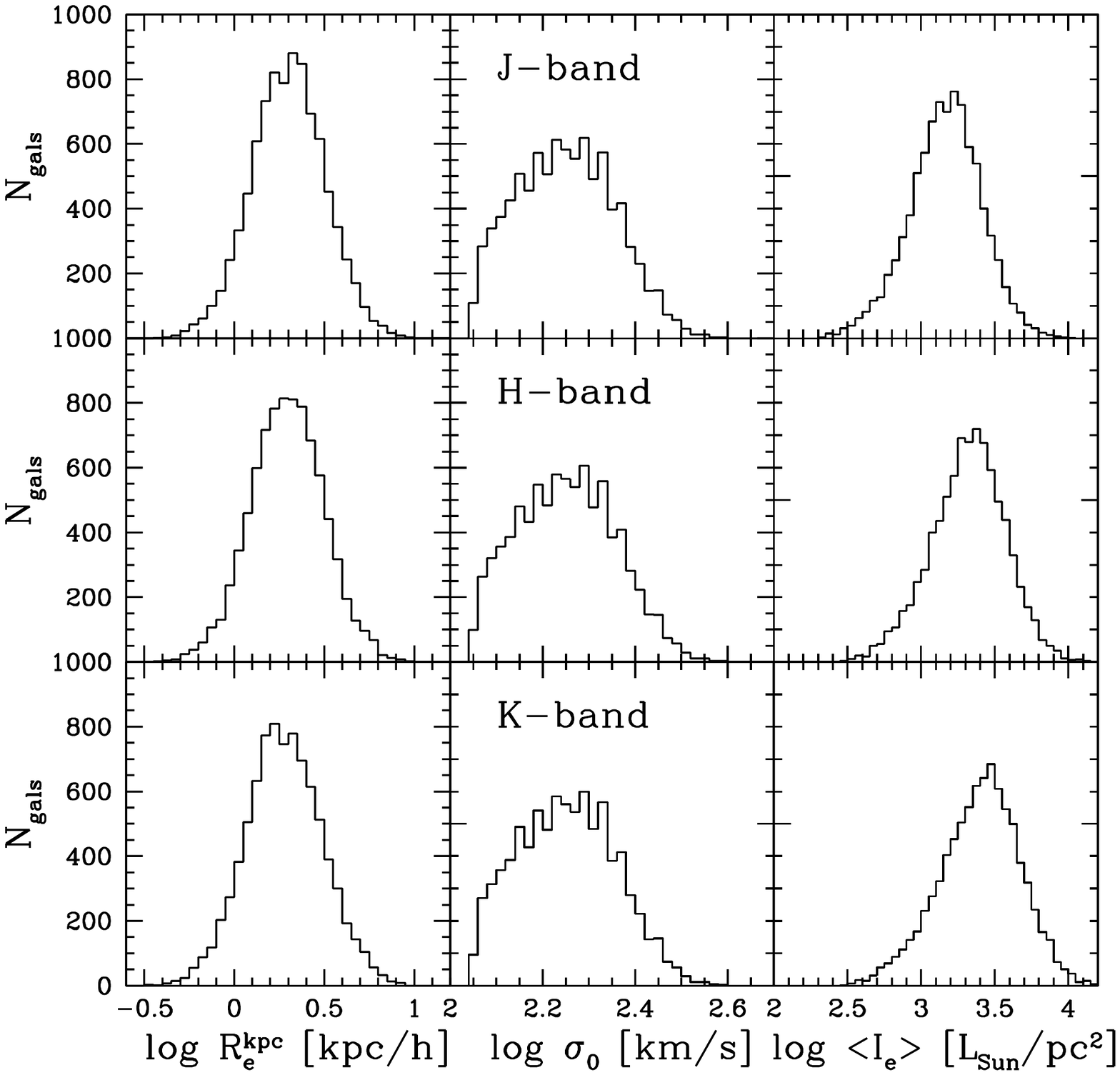} 
\caption{The distribution of the observed FP parameters,
  $\log{R_e^{kpc}}$, \vd\ and \sblum\ for the $J$-, $H$-, and $K$- band
  samples. The top row shows the histograms of the three parameters in
  $J$-band, the middle row in $H$-band, and the bottom row in $K$-band.
  The $\log{R_e^{kpc}}$ and \sblum\ values vary with passband for every
  galaxy, while the \vd\ value for each galaxy remains unchanged.
  However the \vd\ histograms do vary slightly with passband, since the
  $J$-, $H$-, and $K$- band samples include a slightly different mix of
  galaxies. Bin widths on $\log{R_e^{kpc}}$, \vd\ and \sblum\ are 0.05,
  0.02 and 0.05\,dex respectively.}
\label{fig:rsihist}
\end{figure}

\begin{figure}
\centering
\includegraphics[width=0.45\textwidth]{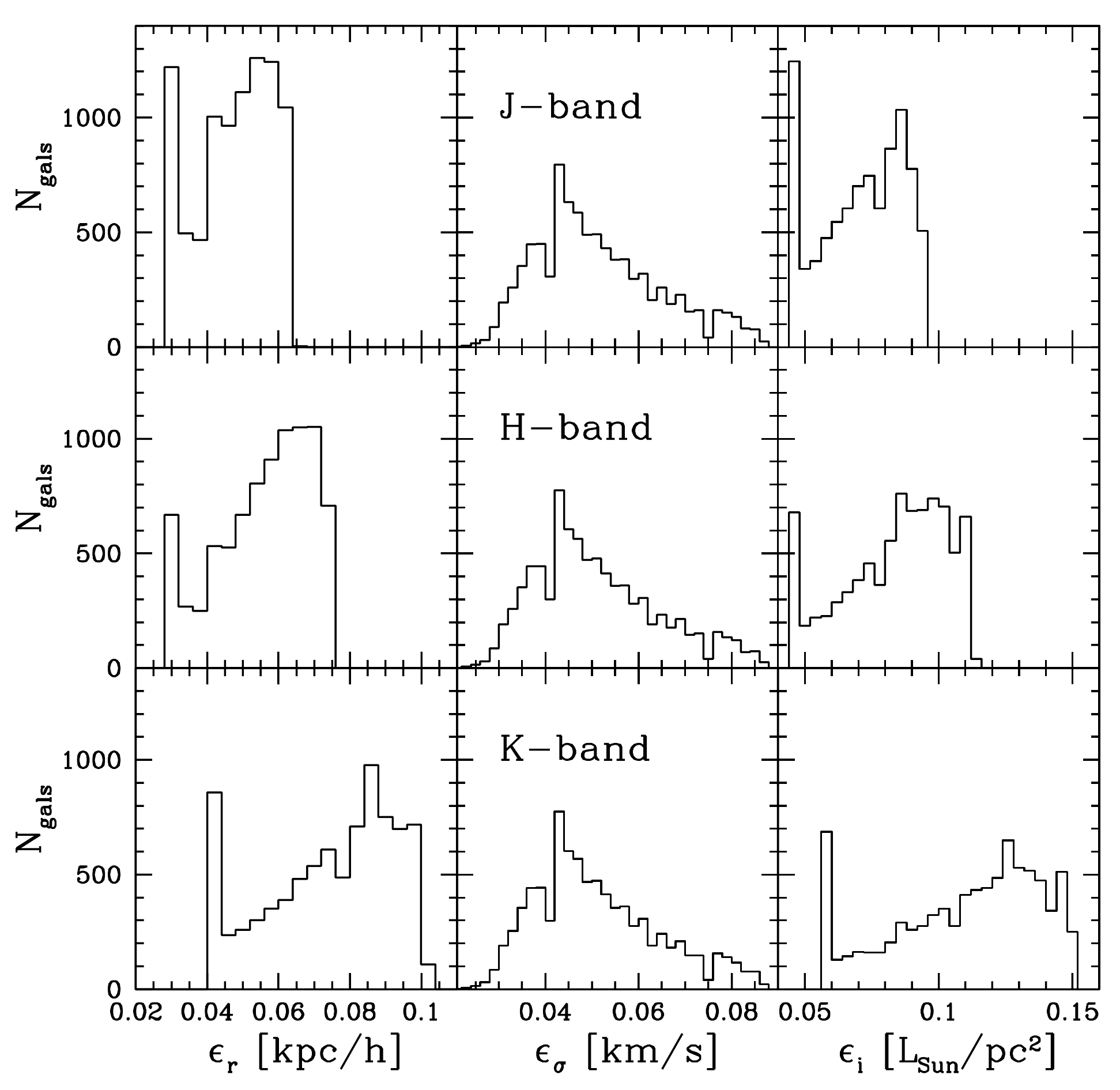} 
\caption{The distribution of the errors on the observed FP parameters
  $\log{R_e^{kpc}}$, \vd\ and \sblum\ for the $J$-, $H$-, and $K$-band
  samples (top to bottom). Bin widths on $\epsilon_{\log{r}}$,
  $\epsilon_{\log{\sigma}}$, and $\epsilon_{\log{i}}$ are 0.004, 0.002
  and 0.004\,dex respectively.}
\label{fig:rsierrhist}
\end{figure}

\begin{figure}
\centering
\includegraphics[width=0.45\textwidth]{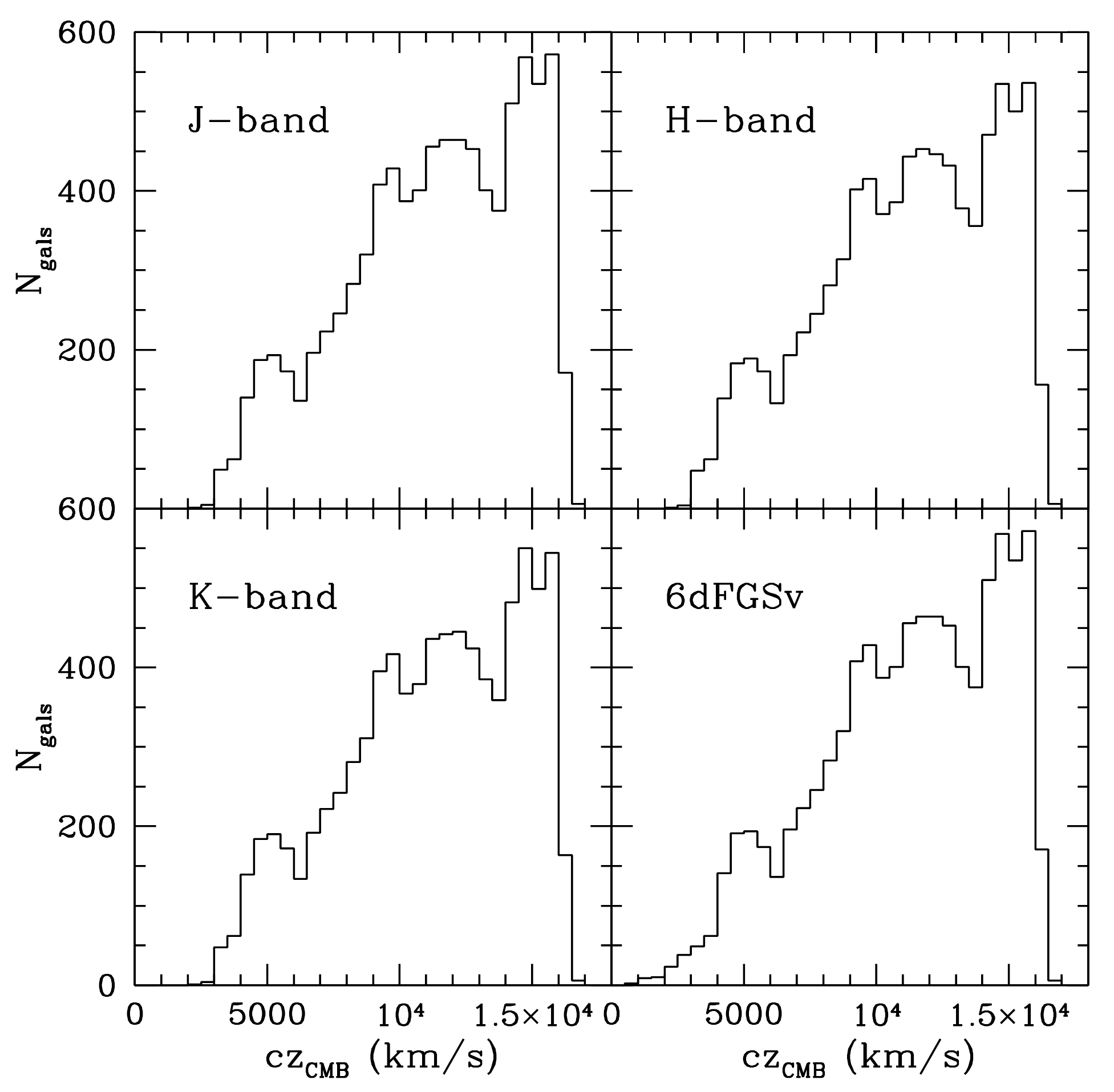}
\caption{The redshift distributions for the $J$-band (upper left),
  $H$-band (upper right), $K$-band (lower left) and the 6dFGSv peculiar
  velocity (lower right) samples. The bins have width 500\kms.}
\label{fig:czhist}
\end{figure}

In Figure~\ref{fig:rsihist} we show histograms of $\log{R_e^{kpc}}$ and
\sblum\ for all three bands, as well as the histogram of \vd. In
Figure~\ref{fig:rsierrhist} we show histograms of the errors on each of
these quantities, while in Figure~\ref{fig:czhist} we show the
histograms of redshifts for the $J$-, $H$-, and $K$-band samples, as
well as the peculiar velocity sample 6dFGSv (in $J$-band).

This is the first public release of the 6dFGS Fundamental Plane peculiar
velocity catalogue (6dFGSv). In previous work (\citealt{magoulas12,
springob12}) we used a preliminary version of this catalogue.
Table~\ref{tab:fpdata} is an updated version of our original catalogue.

The main changes are: (i)~the inclusion of improved velocity dispersion
errors, calculated via a bootstrap technique as described in Section~2.3
(previously these were derived using the TD79 error formula); (ii)~after
visually inspecting the 2MASS postage stamp images and the GALFIT
residual images for all objects, exclusion of 129 galaxies where either
the 2MASS processing removed a substantial part of the target galaxy,
(e.g.\ 2MASXJ08532460--0919124) or the presence of a strong core
asymmetry indicated multiple structures (e.g.\ 2MASXJ13525793--4751401);
(iii)~improved masking of contaminating stars in the GALFIT analysis for
$\sim$100 galaxies; (iv)~the use of updated Galactic extinction values
from \cite{schlafly11} rather than from \cite{schlegel98}; and
(v)~recalculation of the individual galaxy sample codes as a result of the
above changes to the input dataset.

There are no substantive changes to the conclusions reached in our
previous work when this updated 6dFGSv catalogue is used, as (apart from
the new velocity dispersion errors) the changes to the dataset are very
small. In addition, our FP analysis procedure \citep[see][]{magoulas12}
was robust, as outliers were iteratively culled. Further details are
reported in Magoulas et al.\ (2014, {\it in preparation}).

\subsection{6dFGS stellar population parameters}

For completeness we also present here, in Table~\ref{tab:spp_data}, the
stellar population parameters for the 6dFGSv sample that were derived in
\citet{proctor08}.

\begin{table*}
  \caption{6dFGS Stellar Population Catalogue. Only galaxies that meet
    the criteria S/N $\geq$ 9\,\AA$^{-1}$ and $Q_{SP}$ $\leq$ 10 are listed
    here (see \citealt{proctor08}). 
    The columns are:
    (1) source name in 6dFGS catalogue;
    (2) logarithm of the galaxy age in Gyr;
    (3) error in log age, $\epsilon_{age}$;
    (4) Fe abundance, [Fe/H];
    (5) error in [Fe/H], $\epsilon_{Fe/H}$;
    (6) $\alpha$-element over-abundance ratio, [$\alpha/Fe$];
    (7) error in [$\alpha$/Fe], $\epsilon_{\alpha/Fe}$;
    (8) overall metallicity, [Z/H];
    (9) error in [Z/H], $\epsilon_{Z/H}$;
    (10) quality of the stellar population parameter fits, $Q_{SP}$.
    The full version of this table is provided in the online Supporting 
    Information.
  }
\label{tab:spp_data}
\begin{tabular}{crrrrrrrrr}
\hline \hline
\multicolumn{1}{c}{6dFGS ID} & 
\multicolumn{1}{c}{$\log{\rm {age}}$} &
\multicolumn{1}{c}{$\epsilon_{age}$} &
\multicolumn{1}{c}{[Fe/H]} &
\multicolumn{1}{c}{$\epsilon_{Fe/H}$} &
\multicolumn{1}{c}{[$\alpha/Fe$]} &
\multicolumn{1}{c}{$\epsilon_{\alpha/Fe}$} &
\multicolumn{1}{c}{[Z/H]} &
\multicolumn{1}{c}{$\epsilon_{Z/H}$} &
\multicolumn{1}{c}{$Q_{SP}$} \\
\\
\multicolumn{1}{c}{(1)} &  
\multicolumn{1}{c}{(2)} & 
\multicolumn{1}{c}{(3)} &
\multicolumn{1}{c}{(4)} &
\multicolumn{1}{c}{(5)} & 
\multicolumn{1}{c}{(6)} &
\multicolumn{1}{c}{(7)} & 
\multicolumn{1}{c}{(8)} &  
\multicolumn{1}{c}{(9)} &
\multicolumn{1}{c}{(10)} \\
\hline
g0000144-765225 & 0.650 & 0.088 & --0.646 & 0.158 &   0.500 & 0.111 & --0.175 & 0.063 &  5\\
g0000222-013746 & 0.500 & 0.058 &   0.435 & 0.066 & --0.090 & 0.040 &   0.350 & 0.047 &  7\\
g0000235-065610 & 0.475 & 0.075 &   0.237 & 0.088 &   0.120 & 0.038 &   0.350 & 0.067 &  5\\
g0000251-260240 & 0.775 & 0.108 &   0.027 & 0.125 &   0.210 & 0.064 &   0.225 & 0.072 &  6\\
g0000356-014547 & 0.375 & 0.046 & --0.070 & 0.073 &   0.180 & 0.046 &   0.100 & 0.045 &  5\\
g0000358-403432 & 0.500 & 0.107 &   0.005 & 0.151 &   0.180 & 0.070 &   0.175 & 0.099 &  8\\
g0000428-721715 & 1.175 & 0.021 & --0.185 & 0.057 &   0.090 & 0.048 & --0.100 & 0.030 &  4\\
g0000459-815803 & 0.250 & 0.042 & --0.170 & 0.131 &   0.180 & 0.082 &   0.000 & 0.067 &  6\\
g0000523-355037 & 0.700 & 0.083 & --0.471 & 0.113 &   0.500 & 0.073 &   0.000 & 0.054 &  8\\
\hline
\end{tabular}
\end{table*}

These stellar population parameters, which apply to the 6.7\,arcsec
diameter central region of each galaxy covered by a 6dF fibre, are based
on Lick index measurements and the $\chi^2$-fitting procedure of
\citet{proctor02}, which uses all available indices to obtain
simultaneous estimates for log(age), [Fe/H], [Z/H] and [$\alpha$/Fe].
This method is relatively robust against the problems that can afflict
spectral index measurements, particularly when (as for 6dFGS) the
spectra cannot be flux-calibrated, and also, to a certain extent,
against uncertainties in the stellar population models. Here the models
of \citet{korn05} were used, as they provide the required range in both
age and metallicity and include the effects of varying $\alpha$-element
abundances. For the passive galaxies in the FP sample, between 10 and 15
Lick indices ranging from H$\delta$ to Fe5406 were included in the fits.
Some indices in this wavelength range were excluded, however, because
they were poorly fitted by the models: the Ca4455, Mg1 and Mg2 indices
(which are sensitive to flux calibration), and the H$\beta$ and Fe5015
indices (which suffer from low-level emission-line contamination).

For the passive galaxies in the FP sample, the mean reduced $\chi^2$ was
1.08, indicating satisfactory fits given the measured errors in the Lick
indices. For each galaxy, the errors in the stellar population
parameters were estimated as the rms scatter in the values obtained from
50 Monte Carlo simulations of the fitting procedure using the
observational errors in the Lick indices for that galaxy. Excluding low
S/N galaxies with errors in log(age) greater than 0.3\,dex, for which
the stellar population parameters are deemed unreliable, the mean errors
in both log(age) and [Z/H] are approximately 0.15\,dex. These errors
depend on spectral S/N: for galaxies with S/N\,$>$\,16 the mean errors
in both quantities are $\sim$0.1\,dex, while for S/N\,$<$\,16 they are
$\sim$0.2\,dex. The errors in age and metallicity are, as usual,
somewhat correlated. A quality parameter, $Q_{SP}$, was also defined as
the sum of the integerized reduced $\chi^2$ and the number of clipped
indices; galaxy spectra with a $Q_{SP}$ value of 1 have the best quality
fits. Only data from galaxies with S/N per angstrom greater than 9 and
quality parameter of 10 or lower are used in our analysis, and included
in Table~\ref{tab:spp_data} \citet{proctor08} consider all these issues in detail, and
further discussion of the stellar population parameters and their effect
on the FP is provided by \citet{springob12}.

\section{Conclusions}
This paper reports the measurements of Fundamental Plane (FP) parameters
for nearly 9000 galaxies in the 6dFGS peculiar velocity survey (6dFGSv)
and describes how the FP catalogue was constructed. This is the
largest and most homogeneous set of FP measurements for galaxies in the
nearby universe obtained to date.

These data have previously been used to study the effects of stellar
population differences on galaxy scaling relations \citep{proctor08}, to
characterise the near-infrared FP and investigate trends in the FP with
environment and galaxy morphology \citep{magoulas12}, and to map the
changing stellar populations of galaxies as a function of their location
within the FP \citep{springob12}.

In papers now in preparation, this catalogue will be exploited to derive
FP distance estimates for these galaxies, and so determine the peculiar
velocity field throughout the southern hemisphere local volume out to a
redshift of $\sim$16\,000\kms\ \citep{springob14}; this volume samples
many important attractors including the Shapley supercluster.
The velocity field
will, in turn, lead to measurements of the local bulk flow on various
scales and permit comparisons of the observed velocity field with that
predicted from the distribution of galaxies found in redshift surveys
\citep{magoulas14}. The 6dFGS redshifts and peculiar velocities will be
jointly analysed to characterise the statistical properties of the
density and velocity fields captured by the galaxy-galaxy,
galaxy-velocity and velocity-velocity power spectra
\citep{johnson14,koda14}, providing better determinations of
cosmological parameters, such as the redshift space distortion ($\beta$)
and the correlation between galaxies and dark matter ($r_g$), that are
degenerate given only the information provided by redshift surveys.

In the future, the southern 6dFGSv sample will be combined with the
northern SDSS FP sample to recover the local velocity field over about
three-quarters of the sky. Comparisons with new all-sky Tully-Fisher
(TF) peculiar velocity samples, such as the ongoing 2MASS Tully-Fisher
survey \citep[2MTF;][]{masters08, hong13a, hong13b} and the planned
WALLABY/WNSHS surveys \citep{duffy12a,duffy12b}, and with new supernova
surveys \citep[e.g.][]{turnbull12,ma13,rathaus13}, will provide
independent cross-checks and refined measurements. Together, these
peculiar velocity surveys of thousands of galaxies will provide a clear
picture of the motions in the local universe. This will complement the
detailed density maps provided by redshift surveys and lead to tighter
constraints on a wide range of cosmological parameters.

\section*{Acknowledgments}
We acknowledge the efforts of the staff of the Australian Astronomical
Observatory (AAO), who developed the 6dF instrument and carried out the
observations for the survey. This publication makes use of data products
from the Two Micron All Sky Survey, which is a joint project of the
University of Massachusetts and the Infrared Processing and Analysis
Center/California Institute of Technology, funded by the National
Aeronautics and Space Administration and the National Science
Foundation. 
This research has made use of the NASA/IPAC Extragalactic Database
(NED), which is operated by the Jet Propulsion Laboratory, California
Institute of Technology, under contract with the National Aeronautics
and Space Administration.
JRL acknowledges support from STFC via ST/I001573/1.
DHJ acknowledges support from Australian Research Council
Discovery Projects Grant DP-0208876, administered by the Australian
National University. CM and JM acknowledge support from ARC Discovery
Projects Grant DP-1092666. CM was also supported by a scholarship from
the Australian Astronomical Observatory.
We thank the anonymous referee for providing us with constructive
and insightful comments which helped improve this manuscript.
\bibliographystyle{mn2e}
\bibliography{6df_fp}

\end{document}